\UseRawInputEncoding
\documentclass[aps,pra,preprint,superscriptaddress,showpacs,floatfix,floats,showkeys]{revtex4-1}

\usepackage{graphicx}
\usepackage{amssymb}
\usepackage{amsmath}
\usepackage{amsfonts}
\usepackage{booktabs}
\usepackage{dcolumn}
\usepackage{hyperref}
\usepackage{mathrsfs}
\usepackage{multirow}
\usepackage{physics}
\usepackage[caption=false]{subfig}
\usepackage{color}
\definecolor{red}{rgb}{1,0,0}
\newcommand{\ocite}{\cite}
\newcommand{\foreign}[1]{\textit{#1}}

\newcommand{\beq}{\begin{equation}}
\newcommand{\eeq}{\end{equation}}

\newcommand{\DCSD}{$\text{ACCSD}(1{,}\tfrac{3+4}{2})$}
\newcommand{\DCSDt}{$\text{ACCSDt}(1{,}\tfrac{3+4}{2})$}
\newcommand{\DCSDT}{$\text{ACCSDT}(1{,}\tfrac{3+4}{2})$}
\newcommand{\ACCSDX}{$\text{ACCSD}(1{,} 3 \times \tfrac{n_\text{o}}{n_\text{o} + 
		n_\text{u}} + 4 \times \tfrac{n_\text{u}}{n_\text{o} + n_\text{u}})$}
\newcommand{\ACCSDtX}{$\text{ACCSDt}(1{,} 3 \times \tfrac{n_\text{o}}{n_\text{o} 
		+ n_\text{u}} + 4 \times \tfrac{n_\text{u}}{n_\text{o} + n_\text{u}})$}
\newcommand{\ACCSDTX}{$\text{ACCSDT}(1{,} 3 \times \tfrac{n_\text{o}}{n_\text{o} 
		+ n_\text{u}} + 4 \times \tfrac{n_\text{u}}{n_\text{o} + n_\text{u}})$}

\begin{document}

\title{Addressing Strong Correlation by Approximate Coupled-Pair Methods with
Active-Space and Full Treatments of Three-Body Clusters}

\author{Ilias Magoulas}
\affiliation
{Department of Chemistry, Michigan State University, East Lansing, Michigan
48824, USA}

\author{Jun Shen}
\affiliation
{Department of Chemistry, Michigan State University, East Lansing, Michigan
48824, USA}

\author{Piotr Piecuch}
\thanks{Corresponding author}
\email[e-mail: ]{piecuch@chemistry.msu.edu.}
\affiliation
{Department of Chemistry, Michigan State University, East Lansing, Michigan
48824, USA}
\affiliation
{Department of Physics and Astronomy, Michigan State University, East Lansing,
Michigan 48824, USA}

\date{\today}

\begin{abstract}
When the number of strongly correlated electrons becomes larger, the single-reference
coupled-cluster (CC) CCSD, CCSDT, \foreign{etc.}\ hierarchy displays an erratic behavior, while
traditional multi-reference approaches may no longer be applicable due to enormous dimensionalities of 
the underlying model spaces. These difficulties can be alleviated by
the approximate coupled-pair (ACP) theories, in which selected $(T_2)^2$ diagrams
in the CCSD amplitude equations are removed, but there is no generally accepted
and robust way of incorporating connected triply excited ($T_3$) clusters
within the ACP framework. It is also not clear if the
specific combinations of $(T_2)^2$ diagrams that work well for strongly correlated
minimum-basis-set model systems are optimum when larger basis sets are employed.
This study explores these topics by considering a few novel ACP schemes with the
active-space and full treatments of $T_3$ correlations and schemes that scale selected
$(T_2)^2$ diagrams by factors depending on the numbers of occupied and unoccupied orbitals. The
performance of the proposed ACP approaches is illustrated by examining the symmetric
dissociations of the $\text{H}_6$ and $\text{H}_{10}$ rings using basis sets of the
triple- and double-$\zeta$ quality and the $\text{H}_{50}$ linear chain treated with a
minimum basis, for which the conventional CCSD and CCSDT methods fail.
\end{abstract}

\maketitle

\section{Introduction}
\label{sec1}

The size extensive methods based on the exponential wave function ansatz \cite{Hubbard1957a,Hugenholtz1957}
of coupled-cluster (CC) theory \cite{Coester:1958,Coester:1960,cizek1,cizek2,cizek3,cizek4},
\beq
|\Psi \rangle = \exp(T) |\Phi \rangle ,
\label{eq-ccansatz}
\eeq
where
\beq
T = \sum_{n=1}^{N} T_{n}
\label{eq-clusterop}
\eeq
is the cluster operator, $T_{n}$ is the $n$-particle--$n$-hole component of $T$, $N$ is the number of
correlated electrons, and $|\Phi\rangle$ is the reference (\foreign{e.g.}, Hartree--Fock)
determinant defining the Fermi vacuum, have become a \foreign{de facto} standard
for high-accuracy quantum chemistry calculations \cite{paldus-li,Bartlett2007a}.
This is, in significant part,
related to the fact that the conventional single-reference CC hierarchy, including the CC approach
with singles and doubles (CCSD), where $T$ is truncated at
$T_{2}$ \cite{ccsd,ccsd2,ccsdfritz,Piecuch1989}, the CC method
with singles, doubles, and triples (CCSDT), where $T$ is truncated at $T_{3}$ \cite{ccfullt,ccfullt2,ch2-bartlett2},
the CC approach with singles, doubles, triples, and quadruples (CCSDTQ), where $T$ is truncated at $T_{4}$
\cite{ccsdtq0,ccsdtq1,ccsdtq2,ccsdtq3}, \foreign{etc.}, and its extensions to excited states and properties other
than energy through the equation-of-motion \cite{emrich,eomcc1,eomcc3,eomccsdt1,eomccsdt2,eomccsdt3,kallaygauss,hirata}
and linear response \cite{monk,monk2,mukherjee_lrcc,sekino-rjb-1984,lrcc3,lrcc4,jorgensen,kondo-1995,kondo-1996}
formalisms rapidly converge to the exact, full configuration interaction (FCI), limit in weakly correlated systems. 
Higher-order CC methods, such as CCSDT and CCSDTQ,
can also describe multi-reference situations involving smaller numbers of strongly correlated electrons,
encountered, for example, when single and double bond dissociations are examined, allowing one to capture the
relevant many-electron correlation effects
in a conceptually straightforward fashion through particle--hole (p--h)
excitations from a single determinant.

Unfortunately, the conventional CCSD, CCSDT, CCSDTQ, \foreign{etc.}\  hierarchy
may exhibit an erratic behavior and the lack of systematic convergence 
toward the exact, FCI, limit, if the system under consideration is characterized by 
the strong entanglement of larger numbers of electrons, as in the Mott 
metal--insulator transitions \cite{Mott1949,Mott1968,Mott1990}, which can be modeled 
by the Hubbard 
Hamiltonian \cite{Hubbard1963,Hubbard1964a,Hubbard1964b} (see, \foreign{e.g.}, Refs.\ 
\ocite{Vollhardt1984,Imada1998} and references therein) or the linear 
chains, rings, or cubic lattices of the equally spaced hydrogen atoms that change 
from a weakly correlated metallic state at compressed geometries to an 
insulating state with strong correlations in the dissociation region (see, 
\foreign{e.g.}, 
Refs.\ \ocite{Hachmann2006,Bendazzoli2011,Motta2017,Tsuchimochi2009,Sinitskiy2010,%
Kats2013a,Pastorczak2017,evangelista-strong-correlation-2020}).
The analogous challenges apply to the strongly correlated $\pi$-electron 
networks in cyclic polyenes \cite{Pauncz1962a,Pauncz1962b}, as described by the Hubbard 
and Pariser--Parr--Pople (PPP) \cite{Pariser1953,Pariser1953a,Pople1953}
Hamiltonians, which can be used to model
one-dimensional metallic-like systems with Born--von K\'{a}rm\'{a}n periodic 
boundary conditions and a half-filled band \cite{Paldus1984c,Paldus1984b,%
Takahashi1985a,Piecuch1990,Piecuch1991,Paldus1992a,Piecuch1992,Podeszwa2002}.
When the numbers of strongly correlated electrons and open-shell sites from
which these electrons originate become larger, traditional multi-reference methods of the CC 
\cite{paldus-li,Bartlett2007a,lindgren-mukherjee-1987,Piecuch2002b,Lyakh2012a,evangelista-perspective-jcp-2018}
and non-CC \cite{chemrev-2012a,lindh-review-2012,sinha-review-2016} types, which typically build
upon complete active-space self-consistent field (CASSCF) \cite{Ruedenberg1982,Roos1987},
become inapplicable as well (in part, due to rapidly growing dimensionalities
of the underlying multi-configurational reference or model spaces with the numbers of active
electrons and orbitals, which are further complicated by considerable additional
computational costs of determining the remaining dynamical correlation effects needed to obtain a
quantitative description). Even the increasingly popular and undoubtedly promising
substitutes for CASSCF, such as the density-matrix renormalization group (DMRG) approach 
\cite{White1992,White1999,Mitrushenkov2001,Chan2002,Chan2011,Keller2015,Chan2016}
(\foreign{cf.}\ Ref.\ \ocite{dmrg8} for a recent perspective),
FCI Quantum Monte Carlo \cite{Booth2009,Cleland2010,fciqmc-uga-2019,Ghanem2019,ghanem_alavi_fciqmc_2020},
and various selected CI techniques \cite{Whitten1969,Bender1969,Huron1973,Buenker1974,Schriber2016,Schriber2017,%
Tubman2016,Tubman2020,Liu2016,Zhang2020,Holmes2016b,Sharma2017,Li2018,cipsi_1,cipsi_2}, or methods that replace
complete active spaces by their incomplete or multi-layer counterparts (\foreign{cf.}, \foreign{e.g.}, Refs.\
\ocite{ras-1990,Ivanic2003a,Ivanic2003b,gas-2011,gas-2015,las-2019} for selected examples), which allow one
to use significantly larger numbers of active electrons and orbitals compared to CASSCF-based schemes,
begin to wear out when the number of strongly correlated electrons is on the order of 40--50. This is especially
true when one wants to capture the missing dynamical correlations (\foreign{cf.}, \foreign{e.g.}, Refs.\
\ocite{Kurashige2011,Guo2016,Nakatani2017,Sharma2014b,Freitag2017,gaspt2-2016}) and use basis sets much larger than the
minimum one. While there has been a lot of activity directed toward addressing these and related issues,
the challenge of strong correlation remains,
awaiting a satisfactory solution. It is, therefore, desirable to explore various unconventional methodologies
capable of accurately describing weak as well as strong correlation regimes,
especially those that formally belong to the single-reference CC framework,
which is characterized by an ease of implementation and application that
cannot be matched by genuine multi-reference theories. To do this, one has to
understand the origin of the erratic behavior of conventional single-reference CC approaches
in the presence of strong correlations.

The catastrophic failures of the traditional CCSD, CCSDT, CCSDTQ, \foreign{etc.}\ 
hierarchy in all of the aforementioned and similar situations, relevant to condensed 
matter physics, materials science, and the most severe cases of multiple bond 
breaking (\foreign{e.g.}, the celebrated chromium dimer), are related to the 
observation that in order to describe wave functions for $N$ strongly correlated 
electrons one is essentially forced to deal with a FCI-level description of these 
$N$ electrons, which in a conventional CC formulation requires the incorporation of 
virtually all cluster components $T_n$, including $T_N$. Indeed, as shown, for example, in Fig.\ 2 of
Ref.\ \ocite{Degroote2016}, using the 12-site, half-filled attractive pairing 
Hamiltonian with equally spaced levels, or slide 17 of Ref.\ \ocite{Scuseria-lecture},
in which the 10-site Hubbard Hamiltonian with half-filled band is examined,
the higher-order $T_n$ components of the cluster operator, which normally decrease with $n$,
remain large for larger $n$ values approaching $N$ in a strongly correlated regime. This is not
a problem for the single-reference CC ansatz when the number of strongly correlated
electrons $N$ is small (\foreign{e.g.}, 2 in
single bond breaking or 4 in double bond breaking), but becomes a major issue when $N$ is 
larger.

The consequences of the above observations manifest themselves in various, 
sometimes dramatic, ways. For example, one experiences a disastrous behavior of the 
traditional CCSD, CCSDT, CCSDTQ, \foreign{etc.}\ hierarchy, which produces large 
errors, branch point singularities, and unphysical complex solutions in calculations for 
strongly correlated one-dimensional systems modeled by the Hubbard and PPP 
Hamiltonians or $\text{H}_n$ rings, linear chains, and cubic lattices undergoing 
metal--insulator transitions \cite{Paldus1984c,Paldus1984b,Takahashi1985a,
Piecuch1990,Piecuch1991,Paldus1992a,Piecuch1992,Podeszwa2002,Degroote2016,
Bulik2015,Gomez2017a,Gomez2017b,evangelista-strong-correlation-2020}. 
One can also show, using spin-symmetry breaking and restoration arguments, combined 
with the Thouless theorem \cite{Thouless1960,Thouless1961} and a subsequent cluster 
analysis \cite{cizek-paldus-sroubkova-1969}
of projected unrestricted Hartree--Fock (PUHF) wave functions
similar to Refs.\ \ocite{Paldus1984a,Piecuch1996a}, that if we insist on
the wave function ansatz in terms of the $T_2$ cluster component, the 
resulting strongly correlated PUHF state has a non-intuitive polynomial rather than 
the usual exponential form \cite{Qiu2016,Qiu2017,Henderson2017} (\foreign{cf.}, also, 
Refs.\ \ocite{Degroote2016,Gomez2017a,Gomez2017b}). This means that 
in seeking a computationally manageable CC-type solution to a problem of strong correlation
involving the entanglement of many electrons, which would avoid the combinatorial scaling of 
FCI while eliminating failures of the CCSD, CCSDT, CCSDTQ,
\foreign{etc.}\ hierarchy, one has to resign
from the conventional CC treatments in which the cluster operator $T$ is truncated
at a given many-body rank and all terms resulting from the exponential wave function ansatz
are retained.

Among the most interesting solutions in this category are the approaches discussed in Refs.\
\ocite{Degroote2016,Bulik2015,Gomez2017a,Gomez2017b,Qiu2016,Qiu2017,Henderson2017,%
ayers-2013,scuseria1-2014,scuseria2-2014,scuseria3-2014,scuseria4-2014,scuseria1-2016,scuseria2-2016,%
boguslawski-2017,ayers-2017,loos-2021}.
Another promising direction, which is the focus of this study,
is the idea of the approximate coupled-pair
(ACP) approaches \cite{Paldus1984c,Paldus1984b,Takahashi1985a,Piecuch1990,Piecuch1991,
Paldus1992a,Piecuch1992,Podeszwa2002,Piecuch1996a,Paldus1984a,Piecuch1990a,
Adams1981,Jankowski1980,Adams1981b,Chiles1981a,Bachrach1981,Piecuch1995}
and their various more recent reincarnations or modifications, including the 2CC approach and its
$n$CC extensions \cite{Bartlett2006,Musial2007},
the orbital invariant coupled electron pair approximation with an extensive renormalized
triples correction \cite{nooijen-leroy-2006}, the parameterized CCSD methods \cite{Huntington2010}
and their CCSDT-type counterparts \cite{Rishi2019}, and the distinguishable cluster 
approximation with doubles (DCD) or singles and doubles (DCSD) \cite{Kats2013a,
Kats2014,Kats2015,Kats2016,Kats2018} (see, also, 
Refs.\ \ocite{Rishi2016,Rishi2017,Rishi-Perera-Bartlett-2019}) and its DCSD(T) \cite{Kats2016}
and DCSDT \cite{Kats2019,Kats2021,Rishi2019} extensions to connected triples
(see Ref.\ \ocite{Paldus2017} for a review). At the doubles or singles and doubles levels,
the ACP methods have the relatively inexpensive $n_\text{o}^2 n_\text{u}^4$ or ${\mathscr N}^6$
computational costs similar to CCD/CCSD, but by using the appropriately chosen subsets of non-linear
$(T_{2})^{2}$ diagrams of the CCD/CCSD amplitude equations, they greatly improve the performance of
CCD/CCSD in strongly correlated situations, including single and multiple bond dissociations 
\cite{Kats2013a,Huntington2010,Kats2014,Kats2015,Kats2016,Kats2018,Rishi2016,Piecuch1996b}
and, what is particularly intriguing, the low-dimensional metallic-like systems and 
symmetrically stretched hydrogen rings, linear chains, and cubic lattices, where the 
conventional CC treatments completely break down \cite{Kats2013a,Paldus1984c,%
Paldus1984b,Takahashi1985a,Piecuch1990,Piecuch1991,Paldus1992a,Piecuch1992,Podeszwa2002}
(we use the usual notation in which $n_\text{o}$ and $n_\text{u}$
are the numbers of correlated occupied and unoccupied orbitals,
respectively, and ${\mathscr N}$ is a measure of the system size).
As further elaborated on in Section \ref{sec2}, these improvements in the performance of 
conventional single-reference CC approaches are not a coincidence. One can prove that there exist
subsets of CCSD diagrams that result in an exact description of certain strongly 
correlated minimum-basis-set model systems \cite{Piecuch1991,Piecuch1996a,Paldus1984a},
\foreign{i.e.}, the ACP methodologies 
provide a rigorous basis for developing relatively inexpensive CC-like schemes 
for strong correlations (a multi-reference extension of the ACP ideas can
also help genuine multi-reference CC approaches, especially
when multi-determinantal model spaces become inadequate \cite{Piecuch1993a},
but in this study we focus on the single-reference ACP framework).

Having stated all of the above, there remain several open problems that need to 
be addressed before the ACP methods can be routinely applied to realistic 
strongly correlated systems, \foreign{i.e.}, systems involving larger numbers of 
strongly correlated electrons described by \foreign{ab initio} Hamiltonians and 
larger basis sets. One of the main problems is the neglect of connected triply excited
($T_3$) clusters in typical ACP methods. The low-dimensional model systems
with small band gaps, such as the aforementioned cyclic polyenes
near their
strongly correlated limits,
do not suffer from this a lot \cite{Piecuch1990,Paldus1992a},
since their accurate description relies on $T_n$ 
clusters with even values of $n > 2$, but one cannot produce quantitative results in 
the majority of realistic chemistry applications without $T_3$. The previous attempts to
incorporate connected triply excited clusters within the ACP framework using conventional arguments based
on the many-body perturbation theory (MBPT), similar to those exploited in
${\rm CCD[ST]} \equiv {\rm CCD{+}ST(CCD)}$ \cite{Raghavachari1985},
${\rm CCSD[T]} \equiv {\rm CCSD{+}T(CCSD)}$ \cite{Urban1985},
CCSD(T) \cite{Raghavachari1989}, or CCSDT-1 \cite{Lee1984,Lee1984b},
have only been partly successful \cite{Piecuch1990,Paldus1992a,Piecuch1992,Piecuch1996a,Piecuch1995,Kats2016}.
They improved the ACP results in the weakly and moderately
correlated regions of the cyclic polyene models, but did not help in the strongly correlated regime 
\cite{Piecuch1990,Paldus1992a}.
The aforementioned proposals how to include connected triply excited clusters in the
parameterized CCSD and DCSD methods \cite{Kats2016,Kats2019,Kats2021,Rishi2019} and the $n$CC approaches
with $n > 2$, which incorporate $T_{3}$ as well \cite{Bartlett2006,Musial2007},
while being helpful in some cases of bond breaking, have never been applied to strongly correlated
systems involving the entanglement of larger numbers of electrons. Thus, it remains unclear how
to produce a computationally efficient ACP-type procedure that would include the information about
connected triply excited clusters and work well in such situations at the same time. Another open
problem pertains to the fact that the specific combinations of $(T_{2})^{2}$ diagrams that result in the
ACP methods that work well in the strongly correlated regime of the minimum-basis-set model systems,
such as the $\pi$-electron networks of cyclic polyenes, as described by the Hubbard and PPP
Hamiltonians \cite{Paldus1984c,Paldus1984b,Takahashi1985a,Piecuch1990,Piecuch1991,
Paldus1992a,Piecuch1992,Podeszwa2002,Piecuch1996a,Paldus1984a,Piecuch1990a}
(see Section \ref{sec2} for further information),
may not necessarily be optimum when larger basis sets are employed.

We examine both of these topics in the present study. We deal with the problem
of the missing $T_3$ physics by adopting the active-space CC
ideas \cite{eomccsdt1,eomccsdt2,semi0a,semi0b,semi2,semi3,ccsdtq3,semih2o,alex,ghose,%
semi3b,semi4,semi4new,f2bh,eomkkpp,be3,tceijqc,gour1,gour2,gour3,%
piecuch-qtp,jspp-dea-dip-2013,jspp-dea-dip-2014,Ajala2017}
to incorporate the dominant triply excited amplitudes in the ACP methods in a robust, yet 
computationally affordable, manner.
We show that the active-space triples ACP approaches examined in this work, which,
following the naming convention introduced in Refs.\ \ocite{semi4,semi4new}, are
collectively abbreviated as ACCSDt, do not suffer from
the previously observed \cite{Piecuch1990,Paldus1992a} convergence problems resulting
from the use of MBPT-based estimates of $T_3$ contributions within the ACP framework
in a strongly correlated regime. Furthermore, by incorporating the leading triply excited
cluster amplitudes in an iterative manner, the ACCSDt methods developed in this study
allow the $T_{1}$ and $T_{2}$ clusters and, in particular, the
subsets of $(T_{2})^{2}$ contributions
responsible for an accurate description of strong correlations, to relax in the presence
of the dominant $T_3$ amplitudes. The active-space ACCSDt schemes are also characterized
by the systematic convergence toward their ACCSDT parents, in which $T_3$
clusters are treated fully, when the numbers of active occupied and active
unoccupied orbitals used in the ACCSDt calculations increase.
The issue of adjusting the $(T_{2})^{2}$ diagram combinations in the ACP amplitude equations
to the numbers of occupied and unoccupied orbitals used in the calculations is explored
in this study by testing a novel form of the ACP theory, abbreviated as {\ACCSDX}, and
its extensions accounting for $T_3$ correlations, abbreviated as {\ACCSDtX} and {\ACCSDTX},
which utilize the $n_{\rm o}$- and $n_{\rm u}$-dependent scaling factors multiplying the
$(T_{2})^{2}$ diagrams kept in the calculations. At the singles and doubles level and when
$n_{\rm o} = n_{\rm u}$, \foreign{i.e.}, when a minimum basis set is employed, the
{\ACCSDX} scheme reduces to the DCSD approach of Ref.\ \ocite{Kats2013a},
which, in analogy to the closely related ACP-D13 and ACP-D14 methods introduced in Ref.\ \ocite{Piecuch1991},
becomes exact in the strongly correlated limit of cyclic polyenes modeled by the Hubbard
and PPP Hamiltonians. At the same time, the {\ACCSDTX} approach
becomes equivalent to the ACP-D14 method of Ref.\ \ocite{Piecuch1991} augmented with
$T_{1}$ and $T_{3}$ clusters when $n_{\rm o} \ll n_{\rm u}$, which
does, based on our numerical tests, including those discussed in Section \ref{sec3},
improve the {\DCSDT} results, obtained by embedding {\DCSD} = DCSD in CCSDT, in calculations using
larger basis sets. By examining the symmetric dissociations of the
$\text{H}_6$ and $\text{H}_{10}$ rings, as described by basis sets of the
triple- and double-$\zeta$ quality, for which the exact, FCI, calculations are feasible,
and the $\text{H}_{50}$ linear chain treated with a minimum basis, for which the
nearly exact, DMRG, results are available \cite{Hachmann2006},
we show that the active-space {\ACCSDtX} method
and its {\ACCSDTX} parent accurately reproduce
the FCI ($\text{H}_6$ and $\text{H}_{10}$)
and DMRG ($\text{H}_{50}$) energetics, while improving the {\ACCSDX} results
and eliminating catastrophic failures of CCSD and CCSDT in the strongly correlated regions.
Because of the use of basis sets larger than a minimum one in
calculations for the $\text{H}_6$ and $\text{H}_{10}$ ring systems, we also demonstrate
that {\ACCSDtX} and other ACCSDt schemes recover the corresponding ACCSDT results,
including a strongly correlated regime, at the tiny fraction of the computational cost.

\section{Theory and Computational Details}
\label{sec2}

\subsection{Overview of the ACP Schemes}
\label{sec2.1}

\begin{figure*}
\centering
\includegraphics[scale=1]{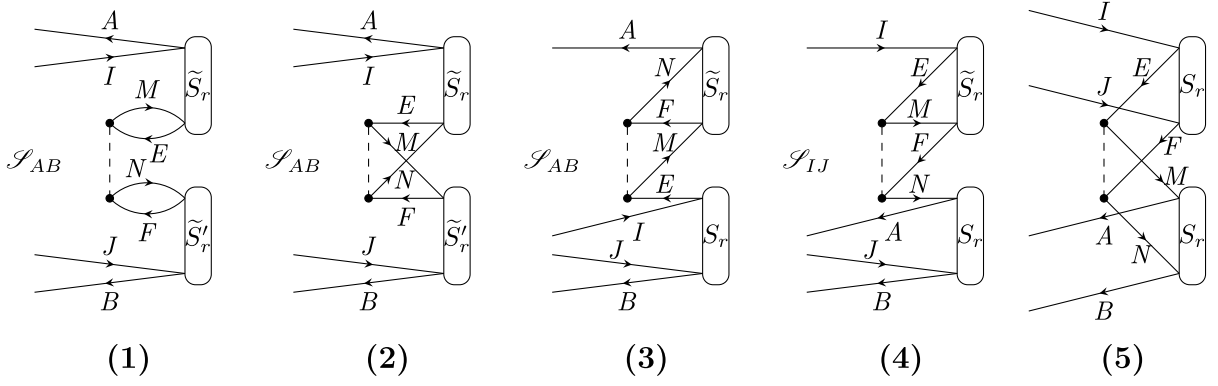}
\caption{\label{figure1}
Goldstone--Brandow orbital diagrams for the $(T_2)^2$
contributions $\Lambda_k^{(2)}(AB,IJ;S_r)$, $k = 1\text{--}5$, to the
CCD or CCSD equations projected on the singlet
pp--hh coupled orthogonally spin-adapted doubly excited 
$\ket{\Phi_{IJ}^{AB}}_{S_r}$ states. The intermediate spin 
quantum number $S_r$ in the definition of the
$\ket{\Phi_{IJ}^{AB}}_{S_r}$ states and the associated doubly 
excited cluster amplitudes $t_{AB}^{IJ}(S_r)$ in the definition of the
orthogonally spin-adapted $T_{2}$ operators,
represented by the Brandow-type, oval-shaped vertices, is 0 or 1. 
The occupied orbital indices $M$ and $N$, the unoccupied orbital indices 
$E$ and $F$, and the intermediate spin quantum numbers 
$\widetilde{S}_r$ and $\widetilde{S}_r^{\prime}$ are summed over.
The ${\mathscr S}_{AB} = 1 + (AB)$ and ${\mathscr S}_{IJ} = 1 + (IJ)$ operators at the diagrams are
index symmetrizers that translate into the symmetrizers or antisymmetrizers,
${\mathscr S}_{AB}(S_{r}) = 1 + (-1)^{S_{r}} (AB)$ and
${\mathscr S}_{IJ}(S_{r}) = 1 + (-1)^{S_{r}} (IJ)$, respectively,
in the resulting algebraic expressions.
}	
\end{figure*}

Historically, a variety of different ways of rationalizing the ACP and related methods using subsets
of non-linear diagrams within a CCD/CCSD framework have been considered (\foreign{cf.}\ Refs.\
\ocite{Kats2013a,Paldus1984c,Piecuch1991,Paldus1992a,Piecuch1996a,Paldus1984a,Piecuch1990a,%
Adams1981,Jankowski1980,Adams1981b,Chiles1981a,Bachrach1981,Bartlett2006,Musial2007,%
nooijen-leroy-2006,Huntington2010,Paldus2017}). Given the objectives of this study, we begin our
discussion with the past numerical observations and mathematical analyses that allow us to understand
the ability of such methods to describe strongly correlated electrons. Let us focus for a moment
on a simpler CCD case, so that we do not have to worry about the less essential, but more
numerous, contributions containing the $T_1$ cluster component. Using the language of
Goldstone--Brandow (Goldstone--Hugenholtz) diagrams, utilized in the derivations 
whenever the orthogonally spin-adapted description \cite{Piecuch1996a,Paldus1984a,%
Piecuch1990a,Cizek1966-tca,Paldus1977b,Paldus1977,Adams1979a,Chiles1981b,lrcc3,%
Piecuch1989,Geertsen1991,Piecuch1992b,Piecuch1994},
important for the understanding of the earliest ACP models,
is desired, one may show that of the five Goldstone--Brandow $(T_2)^2$ diagrams of 
the CCD amplitude equations,
\beq
{}_{S_r}\! \mel{\Phi_{IJ}^{AB}}{[H_N (1 + T_2 + \tfrac{1}{2}T_2^2)]_C}{\Phi} 
= {}_{S_r}\! \mel{\Phi_{IJ}^{AB}}{[H_N (1 + T_2)]_C}{\Phi} + \sum_{k = 1}^{5} 
\Lambda_k^{(2)}(AB,IJ; S_r) = 0,
\label{spin-adapted_ccd}
\eeq
represented in Eq.\ \eqref{spin-adapted_ccd} by the $\Lambda_k^{(2)} 
(AB,IJ; S_r), k= \text{1--5},$ terms and shown in Fig.\ \ref{figure1} as 
diagrams (1)--(5), only two, namely, diagrams (4) and (5), which are separable over the hole line(s),
are needed to eliminate the pole singularities plaguing linearized CCD in situations involving
electronic quasi-degeneracies \cite{Paldus1984c,Paldus1984b,Takahashi1985a,Adams1981,Jankowski1980,Adams1981b}.
Here, we use the notation in which
$H_{N} = H - \langle \Phi | H | \Phi \rangle$ is the Hamiltonian in the normal-ordered
form and subscript $C$ indicates the connected operator product. The
$\ket{\Phi_{IJ}^{AB}}_{S_r}$ states in Eq.\ \eqref{spin-adapted_ccd},
where $I$ and $J$ designate the occupied and
$A$ and $B$ the unoccupied spatial orbitals in the closed-shell reference determinant $|\Phi\rangle$
and $S_r = 0$ or 1 is the intermediate spin quantum number,
represent the singlet particle-particle--hole-hole (pp--hh) coupled orthogonally
spin-adapted doubly excited configuration state functions (CSFs). Using these functions, the $T_{2}$ operator
entering Eq.\ \eqref{spin-adapted_ccd}, represented in Fig.\ \ref{figure1} by
the oval-shaped vertices, is defined as
\beq
T_{2} |\Phi\rangle = \sum_{I \leq J,A \leq B} \sum_{S_r = 0}^{1}
t_{AB}^{IJ}(S_r) \, \ket{\Phi_{IJ}^{AB}}_{S_r}
                   = \frac{1}{4} \sum_{IJAB} \sum_{S_r = 0}^{1}
(N_{IJ}^{AB})^{-2} \,
t_{AB}^{IJ}(S_r) \, \ket{\Phi_{IJ}^{AB}}_{S_r} ,
\label{t2defosa}
\eeq
where $t_{AB}^{IJ}(S_r)$ are the corresponding spin-adapted doubly excited cluster amplitudes
and
\beq
N_{IJ}^{AB} = [(1 + \delta_{IJ}) (1 + \delta_{AB})]^{-1/2} ,
\label{normalization}
\eeq
with $\delta_{IJ}$ and $\delta_{AB}$ designating Kronecker deltas, is the appropriate normalization
factor \cite{Piecuch1996a,Paldus1984a,Piecuch1990a,Paldus1977b,Paldus1977,Adams1979a,Piecuch1989}.
The ACP approach obtained using only the 4th and 5th $(T_2)^2$ contributions in Eq.\ \eqref{spin-adapted_ccd},
$\Lambda_4^{(2)} (AB,IJ; S_r)$ and $\Lambda_5^{(2)} (AB,IJ; S_r)$, respectively, has originally been called
ACP-D45 \cite{Adams1981,Jankowski1980,Adams1981b} or ACCD \cite{Chiles1981a,Bachrach1981}. As shown in Refs.\
\ocite{Paldus1984c,Paldus1984b,Takahashi1985a,Adams1981,Jankowski1980,Adams1981,Adams1981b,%
Chiles1981a,Bachrach1981},
using the aforementioned cyclic polyene models, ${\rm C}_{N}{\rm H}_{N}$, in a $\pi$-electron approximation,
described by the Hubbard and PPP Hamiltonians, where one places $N = 4n + 2$, $n = 1, 2, \ldots ,$
carbon atoms on a ring, and
several \foreign{ab initio} systems, including small hydrogen clusters, beryllium 
atom, and small molecules, the ACP-D45 method is as accurate as CCD in weakly 
correlated cases with no electronic quasi-degeneracies, while representing an 
excellent approximation in strongly correlated, highly degenerate situations, where 
the linearized 
CCD \cite{Paldus1984c,Paldus1984b,Takahashi1985a,Adams1981,Jankowski1980,Adams1981b} 
or even the full CCD \cite{Paldus1984c,Paldus1984b,Takahashi1985a,Piecuch1990,%
Piecuch1991,Paldus1992a,Piecuch1992},
CCD corrected for connected triples via perturbative approximations of the CCD[ST] or CCSD[T] and CCSDT-1 
types \cite{Piecuch1990,Paldus1992a,Piecuch1992}, CCSDT \cite{Podeszwa2002}, and 
CCSDTQ \cite{Podeszwa2002} are plagued with singularities or divergent behavior.
At the time of the initial 
discovery of the ACP-D45 or ACCD scheme, this remarkable behavior was partially 
explained by the mutual cancellation of the contributions arising from the first 
three diagrams in Fig.\ \ref{figure1}, observed 
numerically \cite{Paldus1984c,Adams1981,Jankowski1980,Adams1981b}
and advocated mathematically by considering the limit of non-interacting electron 
pairs \cite{Chiles1981a}. However, this could not explain why the ACP-D45 approach, 
using only two of the five $(T_2)^2$ Goldstone--Brandow diagrams of CCD, works so 
well in the strongly correlated, $\beta = 0$, limit of the cyclic polyene models, 
where CCD completely fails, producing branch point singularities and complex 
solutions for cyclic polyenes with 14 or more carbon sites as $\beta$ approaches
0 from the weakly correlated $\beta \ll 0$ region \cite{Paldus1984c,%
Paldus1984b,Takahashi1985a,Piecuch1990,Piecuch1991,Paldus1992a,Piecuch1992}
($\beta$ is the parameter scaling the one-electron part of the Hubbard or PPP 
Hamiltonians; in the Hubbard Hamiltonian case, $\beta$ is equivalent to $-t/U$, where 
$t$ and $U$ are the parameters controlling the kinetic energy characterizing the 
hopping of electrons between nearest neighbors and on-site electron-electron 
repulsion, respectively). In fact, it was observed that ACP-D45 is exact in the 
strongly correlated, $\beta = 0$, limit of the Hubbard model \cite{Paldus1984c,%
Paldus1984b,Takahashi1985a,Piecuch1990,Piecuch1991,Paldus1992a,Piecuch1992},
while being accurate for the PPP Hamiltonian and other $\beta$ values. Other 
selections of CCD or CCSD $(T_2)^2$ diagrams were considered by several authors in 
recent years \cite{Kats2013a,Bartlett2006,Musial2007,Huntington2010,Kats2014,%
Kats2015,Kats2016,Kats2018,Rishi2016,Rishi2017,Rishi-Perera-Bartlett-2019},
with some choices being similar or even identical to the original ACP approaches 
examined in Refs.\ \ocite{Paldus1984c,
Paldus1984b,Takahashi1985a,Piecuch1990,Piecuch1991,Paldus1992a,Piecuch1992,%
Podeszwa2002,Piecuch1996a,Paldus1984a,Piecuch1990a,Adams1981,Jankowski1980,%
Adams1981b,Chiles1981a,Bachrach1981,Piecuch1995},
seeking diagram combinations that work better than CCD/CCSD in bond breaking 
situations, but none of these recent studies have provided rigorous mathematical arguments
why the ACP methods, such as ACP-D45, can be exact or nearly exact in strongly 
correlated situations of the type of those created by the cyclic polyene models in the
$\beta \approx 0$ region.

The major breakthrough in the understanding of the superb performance of the ACP-D45 
approach in a strongly correlated regime came in 1984 \cite{Paldus1984a}, followed by 
two other key articles published in 1991 \cite{Piecuch1991} and 1996 
\cite{Piecuch1996a}. By performing cluster analysis \cite{cizek-paldus-sroubkova-1969} of the PUHF wave function 
within the orthogonally spin-adapted framework, in which the PUHF state was assumed to be exact,
and using the philosophy of the externally corrected CC methods 
\cite{Piecuch1996a,Paldus1984a,Paldus2017,Paldus1994,Stolarczyk1994,%
Peris1997,Peris1999,Li1997,Li1998,LiPaldus2006,ec-ccsdt-cas,Deustua2018,%
Aroeira2020,chan2021ecCC,ecCC-jctc-2021},
the authors of Refs.\ \ocite{Paldus1984a,Piecuch1996a} demonstrated that the $T_4$ cluster
component extracted from PUHF with the help of the Thouless theorem and read into
the CCD system, Eq.\ \eqref{spin-adapted_ccd}, (Ref.\ \ocite{Paldus1984a}) or its CCSD
extension (Ref.\ \ocite{Piecuch1996a}) as the 
$_{S_r}\! \mel{\Phi_{IJ}^{AB}}{(H_N T_4)_C}{\Phi}$ contribution
cancels out the first three of the five diagrams in Fig.\ \ref{figure1}, while
multiplying the fifth diagram in the equations projected on the singlet pp--hh 
coupled orthogonally spin-adapted doubly excited $\ket{\Phi_{IJ}^{AB}}_{S_r}$ 
CSFs with the intermediate spin $S_r = 1$ by a factor of 9 (in principle, the analogous
$T_3$-containing contributions should have been read into the CCD or CCSD systems too,
but they were not, since $T_3$ extracted from the PUHF
wave function, represented as a CC state relative to the restricted Hartree--Fock (RHF) reference determinant,
vanishes \cite{Paldus1984a,Piecuch1996a}). The PUHF wave function provides
exact energies and cluster amplitudes for the PPP and Hubbard Hamiltonian models of 
cyclic polyenes in the strongly correlated, $\beta = 0$, limit, so the resulting 
$\text{ACP-D}45_9$ theory, abbreviated as ACPQ \cite{Paldus1984a} or 
$\text{ACCD}^\prime$ \cite{Piecuch1996a}, using $\Lambda_4^{(2)}(AB,IJ; S_r) + (2S_r 
+ 1)^2 \Lambda_5^{(2)}(AB,IJ; S_r)$ instead of $\sum_{k = 1}^{5} 
\Lambda_k^{(2)}(AB,IJ; S_r)$ to represent the $(T_2)^2$ contributions within the CCD 
system, Eq.\ \eqref{spin-adapted_ccd}, is mathematically exact in this limit,
in complete agreement with the numerical observations \cite{Paldus1984c,%
Paldus1984b,Takahashi1985a,Piecuch1990,Piecuch1991,Paldus1992a,Piecuch1992,Podeszwa2002}
(the $T_{1}$ contributions can be ignored here, since $T_{1} = 0$ for the cyclic polyene models
described by the PPP and Hubbard Hamiltonians).
The aforementioned factor of 9, which is simply the square of the multiplicity associated with
the intermediate spin value $S_{r} = 1$, was a result of using the orthogonally spin-adapted
formalism; without using this formalism and without the exploitation of the appropriate
many-body and angular momentum diagrammatic techniques
in Refs.\ \ocite{Piecuch1996a,Paldus1984a}, one would not be able to 
see the emergence of $(2S_r + 1)^{2}$ at the $\Lambda_5^{(2)}(AB,IJ; S_r)$
term corresponding to diagram (5) in a transparent manner. The original derivation in Ref.\
\ocite{Paldus1984a} assumed that the PUHF wave function has no singlet-coupled singly excited CSFs
relative to RHF, which is true for the PPP and Hubbard Hamiltonian models of cyclic polyenes
due to symmetry, but not in general, so the $\text{ACP-D}45_9 = \text{ACPQ} = \text{ACCD}^\prime$ 
scheme and its more complete CCSD-level extension, where the $T_1$ clusters are included as well, 
termed $\text{ACCSD}^\prime$ \cite{Piecuch1996a}, were rederived in Ref.\ 
\ocite{Piecuch1996a}. Reference \ocite{Piecuch1996a} also examined the
$\text{CCSDQ}^\prime$ approach and its triples-corrected $\text{CCSDQ}^\prime + 
\text{T}(\text{CCSDQ}^\prime) = \text{CCSDQ}^\prime[\text{T}]$ extension, where one 
does not make any assumptions regarding the accuracy of PUHF and extracts the $T_4$ 
cluster component from a PUHF wave function as is,
showing additional improvements
in some cases,
but our focus here is on the philosophy 
represented by the ACP methods, such as $\text{ACCSD}^\prime$. 
If the above factor of
9 at the fifth diagram in Fig.\ \ref{figure1} for the $S_r = 1$ case is 
ignored (and for the cyclic polyene models, as described by the PPP and Hubbard
Hamiltonians, $\Lambda_5^{(2)}(AB,IJ; S_r)$ is generally small), the 
$\text{ACP-D}45_9 = \text{ACPQ} = \text{ACCD}^\prime$ scheme, derived in Ref.\
\ocite{Paldus1984a} and extended to a singles and doubles level in Ref.\
\ocite{Piecuch1996a}, reduces to the original ACP-D45 or ACCD approach
of Refs.\ \ocite{Adams1981,Jankowski1980,Adams1981b,Chiles1981a,Bachrach1981}.
In fact, in the strongly correlated, $\beta = 0$, limit of cyclic polyenes described by
the Hubbard Hamiltonian, the $\Lambda_5^{(2)}(AB,IJ; S_{r} = 1)$ term vanishes \cite{Piecuch1991},
so that the ACP-D45 approximation becomes exact in this case, as observed numerically \cite{Paldus1984c,%
Paldus1984b,Takahashi1985a,Piecuch1990,Piecuch1991,Paldus1992a,Piecuch1992}.
When the above factor of 9 is included, as in the $\text{ACP-D}45_9 = \text{ACPQ} =
\text{ACCD}^\prime$ or $\text{ACCSD}^\prime$ approaches, one ends up with an exact 
description of the strongly correlated, $\beta = 0$, limit of cyclic polyenes
with $4n+2$ carbons on a ring and described by either the Hubbard or PPP Hamiltonians.
This result is independent of $n$, \foreign{i.e.}, it remains valid in the thermodynamic limit.

There are other selections of the $(T_{2})^{2}$ diagrams entering the CCD or CCSD amplitude
equations that can also be exact in the 
strongly correlated regimes of model Hamiltonians. In Ref.\ \ocite{Piecuch1991}, the 
exactness of $\text{ACP-D}45_9 = \text{ACPQ} = \text{ACCD}^\prime$ in the strongly 
correlated, $\beta = 0,$ limit of the cyclic polyene models described by the
Hubbard and PPP Hamiltonians was re-examined using back
substitution of the exact $T_2$ amplitudes extracted from a PUHF wave function into 
the CCD and $\text{ACP-D}45_9$ systems. By doing so, it was proven that the exact 
$T_2$ amplitudes satisfy the latter system, but not the former one, showing that 
$\text{ACP-D}45_9$ is exact when $\beta = 0$ and CCD is not (as already mentioned, 
CCD behaves erratically in the $\beta = 0$ limit). Encouraged by the usefulness of such an 
analysis, the authors of Ref.\ \ocite{Piecuch1991} searched for other combinations 
of the CCD $(T_2)^2$ diagrams that produce exact results at $\beta = 0,$ discovering 
several possibilities. One of them, using diagrams (1) and (4) in Fig.\ 
\ref{figure1} and 
defining the ACP-D14 approach, is $\Lambda_1^{(2)}(AB,IJ; S_r) + 
\Lambda_4^{(2)}(AB,IJ; S_r)$. As shown in Ref.\ \ocite{Piecuch1991} for the cyclic 
polyene models, which have several symmetries, including the p--h
symmetry, $\Lambda_3^{(2)}(AB,IJ; S_r) = \Lambda_4^{(2)}(AB,IJ; S_r)$, 
\foreign{i.e.}, diagrams (3) and (4) in Fig.\ \ref{figure1} are 
equivalent in this case. Thus, 
one can also propose two other methods, ACP-D13 and ACP-D1(3+4)/2, which correspond 
to the following expressions for the $(T_2)^2$ contributions within a CCD or CCSD 
framework: $\Lambda_1^{(2)}(AB,IJ; S_r) + \Lambda_3^{(2)}(AB,IJ; S_r)$ and 
$\Lambda_1^{(2)}(AB,IJ; S_r) + [\Lambda_3^{(2)}(AB,IJ; S_r) + \Lambda_4^{(2)}(AB,IJ; 
S_r)]/2$, respectively. All three methods, ACP-D13, ACP-D14, and ACP-D1(3+4)/2, are 
exact in the strongly correlated, $\beta = 0,$ limit of cyclic polyenes described by 
the Hubbard and PPP Hamiltonians, \foreign{i.e.}, it is worth considering them all. 
The performance of the ACP-D14 approach in the calculations for cyclic polyenes in 
the entire range of $\beta$ values, from the weakly to the strongly correlated 
regimes and systems as large as $\text{C}_{22} \text{H}_{22}$, was examined in Ref.\ 
\ocite{Piecuch1991}, showing the excellent and non-singular behavior similar to the 
$\text{ACP-D}45_9 = \text{ACPQ} = \text{ACCD}^\prime$ methods tested in Refs.\ 
\ocite{Paldus1984c,Paldus1984b,Takahashi1985a,Piecuch1990,Piecuch1991,Paldus1992a,%
Piecuch1992}.
The calculations using the other two approaches, ACP-D13 and ACP-D1(3+4)/2, were not reported in
Ref.\ \ocite{Piecuch1991}, since they would produce identical results due to the p--h
symmetry intrinsic to the cyclic polyene models. The ACP-D1(3+4)/2 approach,
rationalized by the 1991 analysis in Ref.\ \ocite{Piecuch1991}, is equivalent to the
recently pursued DCD method of Refs.\ \ocite{Kats2013a,Kats2014,Kats2015,Kats2016,Kats2018}.
By averaging diagrams (3) and (4), which are the p--h versions of each other, the
ACP-D1(3+4)/2 = DCD model attempts to reinforce the p--h symmetry, even if there is
none in the Hamiltonian. Time will tell if it is beneficial to do this in
\foreign{ab initio} applications involving strongly correlated situations, that is, 
if using DCD and its CCSD-like extension, termed DCSD \cite{Kats2013a,Kats2014,%
Kats2015,Kats2016,Kats2018},
is an overall better idea than using the older $\text{ACP-D}45_9 = \text{ACPQ} =
\text{ACCD}^\prime$ and $\text{ACCSD}^\prime$ approaches. The performance of the
latter approach in bond breaking situations was examined as early as in
1996 \cite{Piecuch1996b}, but our knowledge of the relative performance of these
different ACP variants, especially in large-scale \foreign{ab initio} studies, is 
still rather limited, although various combinations of $(T_2)^2$ diagrams within CCSD have 
been examined in Refs.\ \ocite{Kats2013a,Huntington2010,Kats2014,Kats2015,Kats2016,%
Kats2018,Rishi2016,Rishi-Perera-Bartlett-2019}, showing encouraging results.

There is no doubt that
numerical tests of the various ACP-type approximations will continue to be helpful, but in order 
for these methods to become successful and widely used in the longer term, especially in the 
examination of problems involving larger numbers of strongly correlated electrons 
that cannot be handled by the existing multi-reference methods, selected CI, or DMRG,
one has to address several issues. The above discussion implies that the
ACP-D13, ACP-D14, ACP-D1(3+4)/2 = DCD, and $\text{ACP-D}45_9 = \text{ACPQ} =
\text{ACCD}^\prime$ approaches, obtained by considering subsets of $(T_2)^2$
diagrams within the CCD system, Eq.\ \eqref{spin-adapted_ccd}, and their extensions 
incorporating $T_1$ clusters are more robust than the traditional CCSD, CCSDT, 
CCSDTQ, \foreign{etc.}\ hierarchy in strongly correlated situations, but the main 
rationale behind their usefulness is based on considering strongly correlated limits 
of highly symmetric, minimum-basis-set, model Hamiltonians. It is not immediately 
obvious that the same combinations of $(T_2)^2$ diagrams remain optimum when larger basis
sets, required by quantitative \foreign{ab initio} quantum chemistry, are employed. More importantly,
as already explained above, the $T_3$ physics is absent in the ACP approaches derived
within a CCD/CCSD framework or its PUHF-driven externally corrected extensions,
in which, as shown, for example, in Refs.\ \ocite{Paldus1984a,Piecuch1996a}, $T_3 = 0$.
In other words, the ACP methods
obtained by selecting and modifying $(T_2)^2$ diagrams within CCD or CCSD,
while capturing strong non-dynamical correlations in a computationally 
manageable fashion, even when the numbers of strongly correlated electrons are 
larger, and providing $T_1$ and $T_2$ clusters that are more accurate than those 
obtained with CCD or CCSD, offer incomplete information about dynamical correlation 
effects, which cannot be accurately described without the connected $T_3$ clusters.
As pointed out in the Introduction, the previously developed ACP schemes corrected
for the effects of connected triply excited clusters using arguments originating from MBPT 
\cite{Piecuch1990a,Piecuch1996a,Piecuch1995} have only had partial success when
examining cyclic polyene models \cite{Piecuch1990,Paldus1992a,Piecuch1992}, whereas the recent
attempts to include $T_3$ correlations in parameterized CCSD and DCSD \cite{Kats2016,Kats2019,Kats2021,Rishi2019}
or via the $n$CC hierarchy \cite{Bartlett2006,Musial2007} have not been applied to strongly correlated
systems involving the entanglement of larger numbers of electrons that interest us in
this study most. It is, therefore, useful to consider alternative ways of handling
connected triply excited clusters within the ACP methodology that might result in
practical computational schemes, while having the potential for working well in a
strongly correlated regime. We discuss such approaches in the next subsection.

\subsection{The Proposed ACP Approaches}
\label{sec2.2}

In searching for the combinations of $(T_2)^2$ diagrams that might potentially improve
the ACP results corrected for $T_3$ correlations when larger basis sets are employed,
it is worth noticing that one could retain the exactness of the ACP approaches
using diagrams (1), (3), and (4) in Fig.\ \ref{figure1} in the strongly correlated limit of cyclic
polyenes, as described by the Hubbard and PPP Hamiltonians, by considering other
combinations of diagrams (3) and (4) than those used in the ACP-D13, ACP-D14, and
ACP-D1(3+4)/2 = DCD methods. One could, in fact, replace the $\sum_{k = 1}^{5}
\Lambda_k^{(2)}(AB,IJ; S_r)$ contribution to the CCD
system, Eq.\ \eqref{spin-adapted_ccd}, originating from the five $(T_2)^2$ diagrams
shown in Fig.\ \ref{figure1}, by
\beq
\Lambda_1^{(2)}(AB,IJ; S_r) + \lambda \, \Lambda_3^{(2)}(AB,IJ; S_r) + (1 - \lambda) \,
\Lambda_4^{(2)}(AB,IJ; S_r)
\label{lambda}
\eeq
with an arbitrary value of $\lambda$, and still be exact in the $\beta = 0$ limit of
the cyclic polyene models. In ACP-D1(3+4)/2 = DCD and its DCSD extension incorporating
$T_1$ clusters, one uses $\lambda = \tfrac{1}{2},$ which is appropriate for
cyclic polyenes, as described by the Hubbard and PPP Hamiltonians that have the p--h 
symmetry, and justified in the case of other strongly correlated systems, such as
the hydrogen clusters examined in this work described by the \foreign{ab initio} 
Hamiltonians, as long as one uses a minimum basis set, for which the p--h symmetry is 
approximately satisfied, but this does not necessarily mean that $\lambda = 
\tfrac{1}{2}$ is the optimum choice for larger basis sets,
particularly when the predominantly dynamical $T_3$ correlations
are included in the calculations. By numerically examining several strongly correlated
systems treated with various basis sets, including the dissociating rings and
linear chains composed of varying numbers of hydrogen atoms, such as those discussed
in Section \ref{sec3}, we have noticed that the ACP-D14 approximation, which uses $\lambda =
0$ in Eq.\ \eqref{lambda}, works better when applied to the $(T_2)^2$ diagrams of the
CCSDT amplitude equations projected on the doubly excited CSFs than its $\lambda = \tfrac{1}{2}$
ACP-D1(3+4)/2 = DCD counterpart when $n_\text{u}$ is greater than $n_\text{o}$. This is especially
true in the $n_\text{u} \gg n_\text{o}$ case, \foreign{i.e.}, when larger basis sets are employed.
This observation suggests that in the case of the ACP approaches using diagrams (1), (3), and (4)
of Fig.\ \ref{figure1} within the CCSDT-type framework, it might be beneficial to scale up
diagram (4) in Eq.\ \eqref{lambda} by decreasing the coefficient $\lambda$, \foreign{i.e.},
by increasing $(1 - \lambda)$ at $\Lambda_4^{(2)}(AB,IJ; S_r)$, when $n_\text{u}$ becomes larger.
The simplest expression for $\lambda$ that allows us to accomplish this objective, while
being invariant with respect to $n_\text{o}$ and $n_\text{u}$ if both of these numbers are
simultaneously scaled by the same factor, is $\lambda = \tfrac{n_\text{o}}{n_\text{o} + n_\text{u}}$.
At the singles and doubles level, the resulting ACP scheme, in which the
$\sum_{k = 1}^{5} \Lambda_k^{(2)}(AB,IJ; S_r)$ contribution to the CCSD equations
projected on the doubly excited CSFs is replaced by Eq.\ \eqref{lambda} with
$\lambda = \tfrac{n_\text{o}}{n_\text{o} + n_\text{u}}$, is referred to as the
{\ACCSDX} method. When $n_\text{o} = n_\text{u}$, the {\ACCSDX} scheme reduces to
the DCSD approach, which is, in view of the above discussion,
a desired behavior. At the same time, {\ACCSDX} becomes
equivalent to the extension of the ACP-D14 approximation to the singles and doubles level when
$n_\text{u} \rightarrow \infty$.

In analogy to the ACP-D13, ACP-D14, and ACP-D1(3+4)/2 approaches
augmented with the $T_1$ clusters within a CCSD framework, abbreviated in this article as
ACCSD(1,3), ACCSD(1,4), and {\DCSD}, respectively, which correspond to setting
$\lambda$ in Eq.\ \eqref{lambda} at 1 [ACCSD(1,3)], 0 [ACCSD(1,4)], and $\tfrac{1}{2}$ [{\DCSD}],
the {\ACCSDX} scheme is exact in the strongly correlated, $\beta = 0$, limit
of cyclic polyenes, ${\rm C}_{4n+2}{\rm H}_{4n+2}$, as described by the Hubbard and PPP Hamiltonians,
independent of the value of $n$. Being equivalent to the DCSD method when $n_{\rm o} = n_{\rm u}$, it is also exact for
two-electron systems or non-interacting electron pairs in a minimum basis set description.
Although, unlike DCSD, the {\ACCSDX}
approach is no longer exact for two-electron systems when $n_{\rm u} > n_{\rm o}$,
the errors relative to FCI for species like ${\rm H}_{2}$ and
${\rm HeH}^{+}$ remain very small, on the order of 1--2 \% of the correlation energy,
even when $n_{\rm u} \gg n_{\rm o}$. Furthermore, as shown in Section \ref{sec3}, using the symmetric
dissociations of the $\text{H}_6$ and $\text{H}_{10}$ rings as examples, the use of the {\ACCSDX}
protocol within a CCSDT framework through the {\ACCSDTX} approach and its active-space {\ACCSDtX}
counterpart, which we discuss next, in applications to strongly correlated systems described by
basis sets larger than a minimum one improves the results compared to the ACCSDT and ACCSDt
calculations that adopt the $(T_{2})^{2}$ diagram selections defining ACCSD(1,3), ACCSD(1,4), and
{\DCSD}. While we will continue examining this theory aspect in the future, the benefits of applying
the {\ACCSDX} approximation within a CCSDT-level description outweigh the loss of
exactness
in calculations for two-electron systems with $n_{\rm u} > n_{\rm o}$, especially when the errors relative to FCI are
as small as mentioned above (and none when $n_{\rm u} = n_{\rm o}$).

Given the above discussion, we now move to the robust ways of incorporating $T_{3}$ correlations
in the ACP schemes in which the $\sum_{k = 1}^{5} \Lambda_k^{(2)}(AB,IJ; S_r)$
contribution to the amplitude equations projected on the doubly excited CSFs is replaced by
Eq.\ \eqref{lambda}. Although our focus is on including the connected triply excited clusters in
the {\ACCSDX} approach corresponding to $\lambda = \tfrac{n_\text{o}}{n_\text{o} + n_\text{u}}$
through the use of the active-space {\ACCSDtX} approximation and its {\ACCSDTX} parent,
we discuss the analogous ACCSDt and ACCSDT extensions of the ACCSD(1,3), ACCSD(1,4), and
{\DCSD} = DCSD methods as well.

We recall that the main idea of all active-space CC methods and their excited-state and open-shell
extensions is that of the selection of the leading higher--than--two-body components of the
cluster and excitation operators, such $T_{3}$ or $T_{3}$ and $T_{4}$,
with the help of a small subset of orbitals around the Fermi level relevant to the quasi-degeneracy
problem of interest \cite{eomccsdt1,eomccsdt2,semi0a,semi0b,semi2,semi3,ccsdtq3,semih2o,alex,ghose,%
semi3b,semi4,semi4new,f2bh,eomkkpp,be3,tceijqc,gour1,gour2,gour3,%
jspp-dea-dip-2013,jspp-dea-dip-2014,Ajala2017} (see Ref.\ \ocite{piecuch-qtp} for a review).
Typically, this is accomplished by partitioning the spin-orbitals used in the calculations
into the core, active occupied, active unoccupied, and virtual subsets and constraining the
spin-orbital indices in the $T_{n}$ cluster amplitudes with $n > 2$ and the corresponding
excitation operators such that one can reproduce the parent CCSDT, CCSDTQ, \foreign{etc.}\ energetics,
including problems characterized by stronger multi-reference correlations, at
the small fraction of the computational costs and with minimum loss of accuracy. In the leading
approach in the active-space CC hierarchy, which was originally developed and implemented in Refs.\
\ocite{semi0b,semi2,ccsdtq3,semih2o,ghose} and which is nowadays abbreviated as CCSDt \cite{semi4,semi4new},
we approximate the cluster operator $T$ by
\beq
T^\text{(CCSDt)} = T_1 + T_2 + \boldsymbol{t_3},
\label{t-definition}
\eeq
where $T_1$ and $T_2$ are the usual one- and two-body components of $T$, treated fully,
and the three-body component $\boldsymbol{t_{3}}$, written in the conventional spin-orbital notation
in which $i, j, k, \ldots$ ($a, b, c, \ldots$) designate the spin-orbitals occupied
(unoccupied) in the reference determinant $|\Phi \rangle$, is defined as
\beq
\boldsymbol{t_3} |\Phi \rangle = \sum_{
\begin{subarray}{c}{
i < j < \mathbf{\underline{k}}}
\\
{\mathbf{\underline{a}} < b < c}
\end{subarray}
}
t_{\mathbf{\underline{a}}bc}^{ij\mathbf{\underline{k}}} \,
|\Phi_{ij\mathbf{\underline{k}}}^{\mathbf{\underline{a}}bc} \rangle .
\label{t3-definition}
\eeq
The underlined bold indices in the triply excited cluster amplitudes
$t_{\mathbf{\underline{a}}bc}^{ij\mathbf{\underline{k}}}$ and the corresponding
triply excited determinants
$|\Phi_{ij\mathbf{\underline{k}}}^{\mathbf{\underline{a}}bc} \rangle$ entering
Eq.\ \eqref{t3-definition} denote active spin-orbitals in the respective
categories ($\mathbf{\underline{k}}$ active occupied and $\mathbf{\underline{a}}$
active unoccupied ones). The singly, doubly, and triply excited amplitudes
(or their spin-adapted counterparts) needed
to determine the cluster operator $T^\text{(CCSDt)}$, Eq.\ \eqref{t-definition},
are obtained in the usual way by projecting the
electronic Schr{\" o}dinger equation, with the CCSDt wave function
$|\Psi^{\rm {(CCSDt)}}\rangle = \exp(T^{\rm (CCSDt)}) |\Phi \rangle$
in it, on the excited Slater determinants (in the spin-adapted case, CSFs)
corresponding to the content of $T^\text{(CCSDt)}$. As in the case of all
single-reference CC theories, the CCSDt energy is obtained by projecting the
Schr{\" o}dinger equation on the reference determinant $|\Phi \rangle$.

The CCSDt methodology has several features that are useful in the context of the
ACP considerations pursued in this study. First, and foremost, it allows us
to bring information about the leading $T_{3}$ correlations, which are absent in
ACCSD(1,3), ACCSD(1,4), {\DCSD} = DCSD, and {\ACCSDX}, in a computationally efficient manner.
Indeed, if $N_\text{o}$ ($< n_{\rm o}$) and $N_\text{u}$ ($\ll n_{\rm u}$) designate the
numbers of the active occupied and active unoccupied orbitals, respectively, the CCSDt
protocol, as briefly summarized above, replaces the expensive computational steps of the
parent CCSDT treatment, which scale as $n_{\rm o}^{3} n_{\rm u}^{5}$, by the much more manageable
$N_\text{o} N_\text{u} n_\text{o}^2 n_\text{u}^4$ operations. At the same time, the CCSDt
calculations reduce the $\sim n_{\rm o}^{3} n_{\rm u}^{3}$ (\foreign{i.e.}, ${\mathscr N}^{6}$-type) storage
requirements associated with the full treatment of triply excited cluster amplitudes $t_{abc}^{ijk}$
to the much less demanding $\sim N_{\rm o} N_{\rm u} n_{\rm o}^{2} n_{\rm u}^{2}$ (${\mathscr N}^{4}$-like)
level. The $N_\text{o} N_\text{u} n_\text{o}^2 n_\text{u}^4$ computational steps of CCSDt,
which are essentially equivalent to the polynomial, ${\mathscr N}^6$-type, steps of CCSD multiplied by a
relatively small prefactor proportional to the number of singles in the active space, are also much less
expensive than typical costs of the CASSCF-based multi-reference CC or CI computations.

Another desirable feature of the CCSDt methodology is its non-perturbative character, which
reduces the risk of introducing divergent behavior in a strongly correlated regime observed
in the earlier ACP computations using MBPT-based treatments of triples \cite{Piecuch1990,Paldus1992a,Piecuch1992}.
The iterative character of CCSDt, which, unlike in the previously explored non-iterative triples corrections
to the ACP and related DCSD approaches \cite{Piecuch1990,Paldus1992a,Piecuch1992,Piecuch1996a,Piecuch1990a,%
Piecuch1995,Kats2016}, allows one to adjust the $T_{1}$ and $T_{2}$ clusters, including the
$(T_{2})^{2}$ diagrams responsible for capturing strong correlations, to the dominant $T_3$ contributions,
so that the relevant dynamical and non-dynamical correlation effects are properly coupled, is a
useful feature too. Last but not least, the definition of the triply excited cluster amplitudes
$t_{\mathbf{\underline{a}}bc}^{ij\mathbf{\underline{k}}}$, Eq.\ \eqref{t3-definition}, adopted
by the CCSDt philosophy guarantees systematic convergence toward the full treatment of triples as
$N_\text{o}$ and $N_\text{u}$ approach $n_{\rm o}$ and $n_{\rm u}$, respectively.

All of this suggests that one should be able to efficiently incorporate the connected triply excited clusters
in the ACP considerations by replacing the CCSDt amplitude equations,
\beq
\langle \Phi_{i}^{a} | [ H_{N} (1 + T_{1} + T_{2} + \tfrac{1}{2} T_{1}^{2}
+ \boldsymbol{t_{3}} + T_{1} T_{2} + \tfrac{1}{6} T_{1}^{3}) ]_{C} | \Phi \rangle = 0 ,
\label{ccsdt-singles}
\eeq
\begin{eqnarray}
\langle \Phi_{ij}^{ab} | [ H_{N} (1 & + & T_{1} + T_{2} + \tfrac{1}{2} T_{1}^{2}
+ \boldsymbol{t_{3}} + T_{1} T_{2} + \tfrac{1}{6} T_{1}^{3}
+ T_{1} \boldsymbol{t_{3}}
+ \tfrac{1}{2} T_{1}^{2} T_{2} + \tfrac{1}{24} T_{1}^{4}) ]_{C} | \Phi \rangle
\nonumber
\\
& + & \sum_{k=1}^{5} \Lambda_{k}^{(2)}(ab,ij) = 0,
\label{ccsdt-doubles}
\end{eqnarray}
\begin{eqnarray}
\langle \Phi_{ij\mathbf{\underline{k}}}^{\mathbf{\underline{a}}bc} |
[H_{N} (T_{2} & + & \boldsymbol{t_{3}} + T_{1} T_{2}
+ T_{1} \boldsymbol{t_{3}} + \tfrac{1}{2} T_{2}^{2} + \tfrac{1}{2} T_{1}^{2} T_{2}
\nonumber
\\
& + & T_{2} \boldsymbol{t_{3}} + \tfrac{1}{2} T_{1}^{2} \boldsymbol{t_{3}} + \tfrac{1}{2} T_{1} T_{2}^{2}
+ \tfrac{1}{6} T_{1}^{3} T_{2}) ]_{C} | \Phi \rangle = 0,
\label{ccsdt-triples}
\end{eqnarray}
where $|\Phi_{i}^{a} \rangle$ and $|\Phi_{ij}^{ab} \rangle$ are the singly and doubly excited determinants,
$|\Phi_{ij\mathbf{\underline{k}}}^{\mathbf{\underline{a}}bc} \rangle$ are the selected triply excited determinants
entering the definition of $\boldsymbol{t_{3}}$, Eq.\ \eqref{t3-definition}, and $\Lambda_{k}^{(2)}(ab,ij)$,
$k = 1\text{--}5$, are the five Goldstone--Hugenholtz diagrammatic contributions to the
$\langle \Phi_{ij}^{ab} | ( H_{N} \tfrac{1}{2} T_{2}^{2})_{C} | \Phi \rangle$ term in the equations projected
on the doubly excited determinants, by their ACCSDt counterparts, in which we replace Eq.\ \eqref{ccsdt-doubles}
in the above system by
\begin{eqnarray}
\langle \Phi_{ij}^{ab} | [ H_{N} (1 & + & T_{1} + T_{2} + \tfrac{1}{2} T_{1}^{2}
+ \boldsymbol{t_{3}} + T_{1} T_{2} + \tfrac{1}{6} T_{1}^{3}
+ T_{1} \boldsymbol{t_{3}}
+ \tfrac{1}{2} T_{1}^{2} T_{2} + \tfrac{1}{24} T_{1}^{4}) ]_{C} | \Phi \rangle
\nonumber
\\
& + & \Lambda_{1}^{(2)}(ab,ij) + \lambda \, \Lambda_{3}^{(2)}(ab,ij) + (1 - \lambda) \, \Lambda_{4}^{(2)}(ab,ij) = 0 .
\label{accsdt-doubles}
\end{eqnarray}
As already alluded to above, in this study we focus on the {\ACCSDtX} approach, in which
$\lambda$ in Eq.\ \eqref{accsdt-doubles} is set at $\tfrac{n_\text{o}}{n_\text{o} + n_\text{u}}$,
and its performance in the strongly correlated situations created by the $\text{H}_6$ and $\text{H}_{10}$ rings
and the $\text{H}_{50}$ linear chain, for which the exact, FCI, or nearly exact, DMRG, data are
available, although the closely related ACCSDt(1,3), ACCSDt(1,4), and {\DCSDt} schemes corresponding to
$\lambda = 1$, 0, and $\tfrac{1}{2}$, respectively, are considered in our calculations  as well.
Along with the {\ACCSDtX}, ACCSDt(1,3), ACCSDt(1,4), and {\DCSDt} methods, we examine
their {\ACCSDTX}, ACCSDT(1,3), ACCSDT(1,4), and {\DCSDT} parents, which are obtained by replacing
$\boldsymbol{t_{3}}$ in Eqs.\ \eqref{ccsdt-singles}, \eqref{ccsdt-triples}, and \eqref{accsdt-doubles}
by its $T_{3}$ counterpart, \foreign{i.e.}, by making all orbitals used to define the triply excited amplitudes
and determinants active. Unlike in Refs.\ \ocite{Bartlett2006,Musial2007,Rishi2019,Kats2019,Kats2021},
where the authors were primarily interested in retaining
the exactness for three-electron systems, we do not attempt to alter or simplify the amplitude equations
corresponding to the projections on the triply excited determinants in our ACCSDt and ACCSDT schemes.
Our main interest in this exploratory study is the examination of a strongly correlated regime
of the type of metal--insulator transitions modeled by the dissociating hydrogen rings and linear chains,
and model systems that were used in the past to advocate diagram
cancellations in similar situations, including the aforementioned cyclic polyenes in the Hubbard and PPP Hamiltonian
description, do not provide any specific guidance how to handle the connected triply excited clusters.
Thus, as shown in Eq.\ \eqref{ccsdt-triples}, we keep all terms
in the CC equations projected on the triply excited determinants resulting from the use of the exponential
wave function ansatz in which the cluster operator $T$ is truncated at the three-body component, and
then, whenever
possible, \foreign{i.e.}, when the basis set used in the calculations is larger than a minimum one,
reduce the computational costs associated with the full treatment of $T_{3}$ by constraining
the indices in the $t_{abc}^{ijk}$ amplitudes following the CCSDt recipe defined by Eq.\ \eqref{t3-definition}.


\subsection{Computational Details}
\label{sec2.3}

All of the triples-corrected ACP methods discussed in Section \ref{sec2.2}, which emerge from replacing
Eq.\ \eqref{ccsdt-doubles} in the CCSDt/CCSDT system by Eq.\ \eqref{accsdt-doubles} with the appropriate
$\lambda$ values, including {\ACCSDtX}, ACCSDt(1,3), ACCSDt(1,4), and {\DCSDt}, where the connected triply
excited clusters are treated using active orbitals, and their respective {\ACCSDTX}, ACCSDT(1,3), ACCSDT(1,4),
and {\DCSDT} parents, in which $T_{3}$ clusters are treated fully, have been implemented in our local version
of the GAMESS package \cite{gamess,gamess2,gamess3}. The resulting codes, which work
with the RHF as well as restricted open-shell Hartree--Fock references, and the analogous computer
programs that enable the {\ACCSDX}, ACCSD(1,3), ACCSD(1,4), and {\DCSD} = DCSD calculations,
in which the connected triply excited clusters are neglected, have been generated by making suitable modifications
in our previously developed \cite{jspp-chemphys2012,jspp-jcp2012,jspp-jctc2012,Bauman2017} CCSD, CCSDt, and
CCSDT GAMESS routines. As in the case of other CC methodologies pursued by our group in the past
(\foreign{cf.}\ Section III.B in Ref.\ \ocite{gamess3} for the relevant information),
our plan is to make all of the above ACP schemes, with and without connected triples,
available in the official GAMESS distribution.

In order to test the performance of the ACP methods implemented in this work, especially the benefits
offered by the active-space {\ACCSDtX} scheme and its {\ACCSDTX} parent compared to the remaining
ACCSDt and ACCSDT approaches discussed in Section \ref{sec2.2} and the underlying ACCSD approximations
in a strongly correlated regime, we carried out a series of calculations for the symmetric dissociations of the
$\text{H}_6$ and $\text{H}_{10}$ rings and the $\text{H}_{50}$ linear chain, for which the conventional
CCSD, CCSDt, and CCSDT methods fail as the H--H distances simultaneously increase. In the case of the
$D_\text{6h}$-symmetric $\text{H}_6$ and $D_\text{10h}$-symmetric $\text{H}_{10}$
ring systems, we employed the largest basis sets that allowed us to
perform the exact, FCI, computations using the determinantal FCI code \cite{Ivanic2001,Ivanic2003a,Ivanic2003b}
available in GAMESS. Those were the polarized valence correlation-consistent basis set of the triple-$\zeta$
quality using spherical $d$ functions, commonly abbreviated as cc-pVTZ \cite{Dunning1989}, for $\text{H}_6$,
which results in the many-electron Hilbert space spanned by $\sim 10^{9}$ singlet CSFs,
and the double-$\zeta$ (DZ) basis of Refs.\ \ocite{Dunning1970,Dunning1977} in the case of $\text{H}_{10}$,
giving rise to a FCI problem on the order of $10^{7}$ singlet CSFs (we could not use basis sets of the
triple-$\zeta$ quality or a DZ basis augmented with polarization functions in the latter case,
since the resulting Hamiltonian diagonalizations using the FCI algorithms
available in GAMESS turned out to be prohibitively expensive for us).
In addition to illustrating the computational efficiency of our active-space ACCSDt schemes compared to
their ACCSDT counterparts, the use of basis sets larger than a minimum one for the $\text{H}_6$ and $\text{H}_{10}$
systems allows us to demonstrate that scaling diagrams (3) and (4) in {\ACCSDtX} and {\ACCSDTX} with factors
depending on the numbers of occupied and unoccupied orbitals involved in the calculations can benefit
the resulting energetics. When considering the symmetric dissociation of the equidistant $\text{H}_{50}$ linear chain,
we had to proceed differently. In this case, the exact Hamiltonian diagonalization becomes prohibitively expensive
even when a minimum basis set is employed (when $n_{\rm o} = n_{\rm u} = 25$, one already needs $\sim 10^{27}$ singlet
CSFs to define the corresponding many-electron Hilbert space). Thus, in applying the ACP approaches
examined in this study to the $\text{H}_{50}$ linear chain,
we relied on the nearly exact results reported for this system in Ref.\ \ocite{Hachmann2006}, which were
obtained with the DMRG algorithm abbreviated as LDMRG(500) and the STO-6G minimum basis set \cite{Hehre1969}.
All of the CC (CCSD, CCSDt, and CCSDT) and ACP (ACCSD, ACCSDt, and ACCSDT) calculations reported in this 
article employed the RHF determinant as a reference.

In line with the nature of strong correlations created by the dissociations of the above hydrogen clusters,
the active orbitals used in the ACCSDt and CCSDt calculations consisted of the molecular orbitals (MOs)
that correlate with the 1$s$ shells of the hydrogen atoms. This means that the ACCSDt and CCSDt computations
for the $\text{H}_6$ ring were performed using three active occupied and three active unoccupied
orbitals ($N_{\rm o} = N_{\rm u} = 3$), whereas the analogous calculations for the $\text{H}_{10}$
system used five active occupied and five active unoccupied MOs ($N_{\rm o} = N_{\rm u} = 5$).
Because of the use of the cc-pVTZ basis for $\text{H}_6$ and the DZ basis for $\text{H}_{10}$,
\foreign{i.e.}, basis sets that are considerably larger than a minimum one,
the numbers of active unoccupied orbitals used in the ACCSDt and CCSDt computations for the
dissociating $\text{H}_6$ and $\text{H}_{10}$ rings, especially for $\text{H}_6$,
were much smaller than the numbers of all unoccupied MOs characterizing these systems
(81 and 15, respectively).
In the case of the $\text{H}_{50}$ linear chain, where, to be consistent with Ref.\ \ocite{Hachmann2006}
that provided the LDMRG(500) reference data, we had to use the STO-6G minimum
basis set, the only meaningful active space is that incorporating all 25 occupied and 
all 25 unoccupied orbitals. Thus, our ACCSDt and CCSDt results for this system
are equivalent to those obtained in the respective ACCSDT and CCSDT calculations. Furthermore,
since $n_{\rm o} = n_{\rm u}$ in this case and the differences between the {\ACCSDTX} = {\DCSDT},
ACCSDT(1,3), and ACCSDT(1,4) energies for the $\text{H}_{50}$ linear chain described by the
STO-6G basis set are rather small (for each of the studied geometries,
less than 1\% of the correlation energy),
in reporting our results for this system we focus on the {\DCSDT} calculations
and the associated {\DCSD} = DCSD, CCSD, and CCSDT data.

All of the calculations for the symmetric dissociations of the $\text{H}_6$ and $\text{H}_{10}$
rings reported in this article employed the following grid of internuclear separations between
the neighboring hydrogen atoms, denoted as $R_\text{H--H}$, to determine the corresponding potential
energy curves (PECs): 0.6, 0.7, 0.8, 0.9, 1.0, 1.1, 1.2, 1.3, 1.4,
1.5, 1.6, 1.7, 1.8, 1.9, 2.0, 2.1, 2.2, 2.3, 2.4, and 2.5 {\AA}. The PECs
characterizing the symmetric dissociation of the
$\text{H}_{50}$ linear chain were calculated at the geometries defined by the
$R_\text{H--H}$ values used in Ref.\
\ocite{Hachmann2006}, namely, 1.0, 1.2, 1.4, 1.6, 1.8, 2.0, 2.4, 2.8, 3.2, and 3.6 bohr
(the authors of Ref.\ \ocite{Hachmann2006} considered one additional H--H distance of 4.2 bohr,
but we run into difficulties with converging our {\DCSDT} calculations in this case,
so the largest H--H separation reported in this work is 3.6 bohr).

\section{Numerical Results}
\label{sec3}

\subsection{Symmetric Dissociation of the \mbox{\boldmath{${\text{H}}_{6}$}} and
\mbox{\boldmath{${\text{H}}_{10}$}} Rings}
\label{sec3.1}

\begin{figure*}
	\centering
	\includegraphics[scale=1]{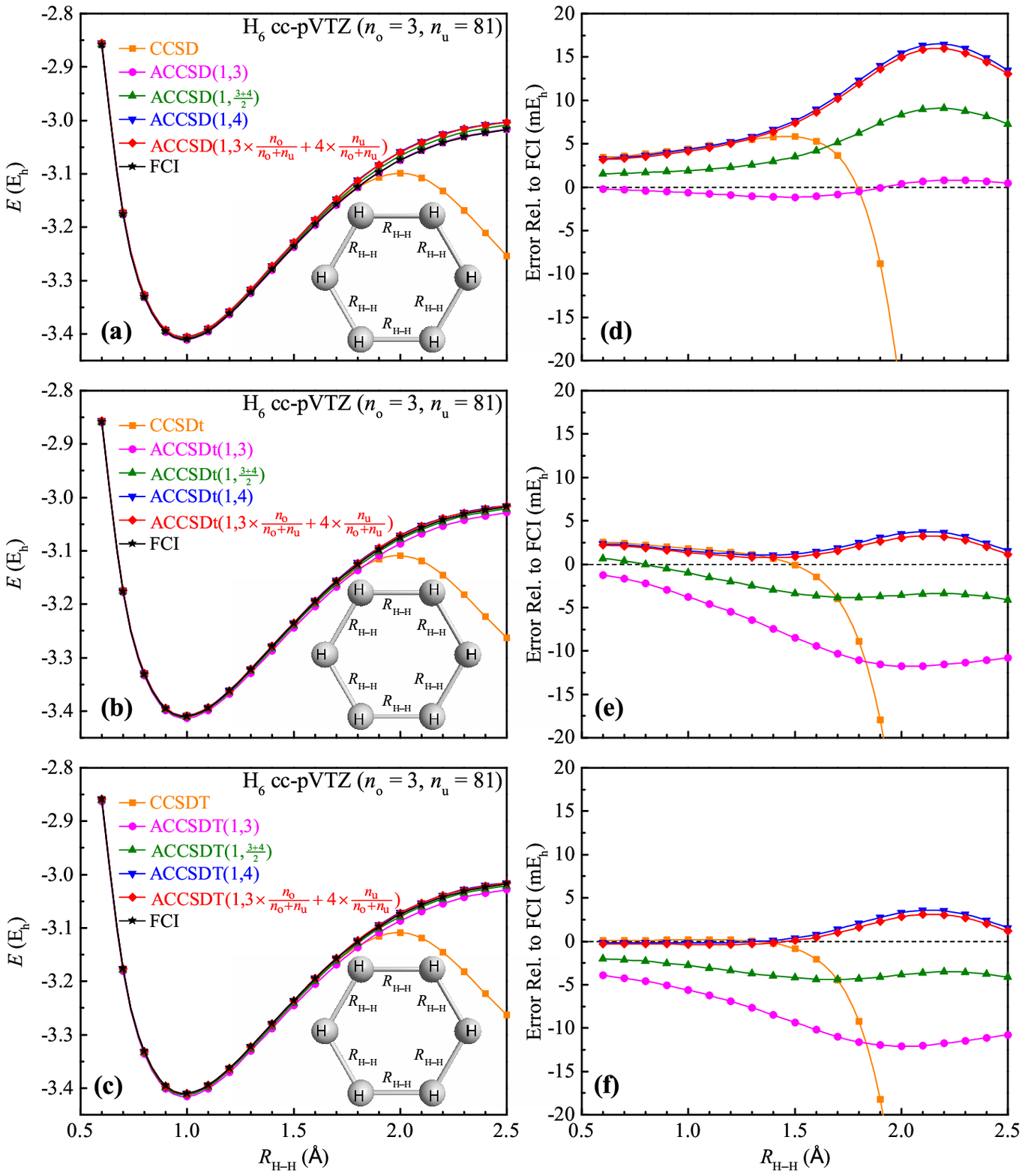}
	\caption{\label{figure2}
		Ground-state PECs [panels (a)--(c)] and errors relative 
		to FCI [panels (d)--(f)] for the symmetric dissociation 
		of the $\text{H}_{6}$ ring resulting from the CCSD and various ACCSD
		[panels (a) and (d)], CCSDt and various ACCSDt 
		[panels (b) and (e)], and CCSDT and various ACCSDT
		[panels (c) and (f)] calculations using the cc-pVTZ basis set.
                The CCSDt and ACCSDt
		approaches employed a minimum active space consisting of three occupied and three
                lowest-energy unoccupied MOs that correlate with the 1$s$ shells
		of the individual hydrogen atoms. The FCI PEC included in panels 
		(a)--(c) is shown to facilitate comparisons.
	}
\end{figure*}

We begin our discussion of the numerical results obtained in this work by examining the $D_\text{6h}$-symmetric 
dissociation of the six-membered hydrogen ring, as described by the cc-pVTZ 
basis set. The information about the ground-state PECs of the $\text{H}_{6}$/cc-pVTZ system obtained in
the conventional CCSD, CCSDt, and CCSDT calculations, their ACCSD, ACCSDt, and ACCSDT
counterparts, and the exact FCI diagonalizations is summarized in Tables
\ref{ACP_H6_doubles}--\ref{ACP_H6_fullt} and Fig.\ \ref{figure2}. Table \ref{ACP_H6_doubles}
reports the results of the computations performed with the various CCSD-type levels,
including CCSD, ACCSD(1,3), {\DCSD} = DCSD, {\ACCSDX}, and ACCSD(1,4).
The analogous results obtained with the CCSDt-type methods, which describe the effects
of $T_{3}$ correlations with the help of active orbitals, including
CCSDt, ACCSDt(1,3), {\DCSDt}, {\ACCSDtX}, and ACCSDt(1,4), are shown in
Table \ref{ACP_H6_active}. Table \ref{ACP_H6_fullt} summarizes the calculations
carried out with the CCSDT, ACCSDT(1,3), {\DCSDT}, {\ACCSDTX}, and ACCSDT(1,4),
approaches, in which $T_{3}$ clusters are treated fully. As already alluded to above,
because of the use of the cc-pVTZ basis set in our calculations for the hexagonal $\text{H}_{6}$
system, the dimension of the corresponding many-electron Hilbert space is on the order of $10^{9}$
(in our FCI calculations employing the determinantal GAMESS routines that permit the use of Abelian symmetries,
the number of determinants relevant to the exact ground-state problem, \foreign{i.e.}, the $S_{z} = 0$
determinants belonging to the $A_{g}$ irreducible representation of the largest Abelian subgroup of $D_\text{6h}$,
which is $D_{2 \rm h}$, was 1,139,812,264).
The {\DCSD} = DCSD calculations for the symmetric dissociation of
the $\text{H}_{6}$ ring system were also reported in Ref.\ \ocite{Kats2013a}, but, unlike in the
present study, the authors of Ref.\ \ocite{Kats2013a} used a minimum basis set, which results in
the many-electron Hilbert space 7 orders of magnitude smaller than that generated with cc-pVTZ, and
did not consider the effects of $T_{3}$ correlations on the resulting PEC. It is, therefore, interesting
to examine how the use of a much larger cc-pVTZ basis and the incorporation of the connected
triply excited clusters in the calculations affects the observations made in Ref.\ \ocite{Kats2013a}.
It is also worth investigating if the replacement of the {\DCSDT} scheme and its active-space
{\DCSDt} counterpart, which build upon DCSD, by the {\ACCSDTX} and {\ACCSDtX} approaches
proposed in this work provides improvements over the results of the analogous {\DCSDT} and {\DCSDt}
calculations.

As shown in Tables \ref{ACP_H6_doubles}--\ref{ACP_H6_fullt} and Fig.\ \ref{figure2}, the
conventional CCSD, CCSDt, and CCSDT methods applied to the symmetric dissociation of
the $\text{H}_{6}$/cc-pVTZ ring system fail when the strongly correlated region of
larger H--H separations is considered. In fact, they break down rather quickly as
the $R_\text{H--H}$ values increase. This can be illustrated by the very large negative
errors relative to FCI resulting from the CCSD, CCSDt, and CCSDT calculations in the
$R_\text{H--H} \geq 2.0$ {\AA} region, which grow, in absolute value, from about
25--35 $\text{mE}_\text{h}$ at $R_\text{H--H} = 2.0$ {\AA} to gargantuan 237--246 $\text{mE}_\text{h}$
when the distance between the neighboring H--H atoms in the hexagonal $\text{H}_{6}$
system reaches 2.5 {\AA}. All three CC approaches produce completely erratic, similarly
shaped, PECs, with an unphysical hump at the intermediate stretches of the H--H
bonds and a well-displayed downhill behavior as the $R_\text{H--H}$ values approach the
strongly correlated asymptotic region. As one might have anticipated based on the examination
of strongly correlated model systems \cite{Podeszwa2002,Degroote2016,Scuseria-lecture} (\foreign{cf.}, also, Refs.\
\ocite{Piecuch1990,Paldus1992a}), the explicit inclusion of the connected triply excited
clusters through the full CCSDT treatment and its active-space CCSDt counterpart offers no help
in this regard. The CCSDt and CCSDT approaches improve the CCSD results in the weakly correlated
equilibrium region, reducing the 4.424 $\text{mE}_\text{h}$ error relative to FCI obtained with
CCSD at $R_\text{H--H} = 1.0$ {\AA} to 1.804 $\text{mE}_\text{h}$ and 0.160 $\text{mE}_\text{h}$,
respectively, but they are as inaccurate as CCSD (or even less accurate) when the H--H distances
become larger. The only benefit of using CCSDt is a major reduction in the computational
timings compared to CCSDT, from 98 s per iteration in the latter case to 6 s per iteration
in the case of the former method (timings obtained using a single core of the Precision 7920 system
from Dell equipped with 10-core Intel Xeon Silver 4114 2.2 GHz processor boards), with minimum
loss of accuracy, especially at larger H--H separations, but the fact that the CCSDt approach
faithfully reproduces the CCSDT PEC at the small fraction of the effort is of no help
here, since both PECs are qualitatively incorrect.

A quick inspection of Fig.\ \ref{figure2} reveals that all ACP methods examined in this study, 
without and with the connected triples, provide qualitatively correct PECs for the
$D_\text{6h}$-symmetric dissociation of the $\text{H}_{6}$/cc-pVTZ ring. As demonstrated
in Tables \ref{ACP_H6_doubles}--\ref{ACP_H6_fullt} and Fig.\ \ref{figure2}, they eliminate
the erratic behavior of the conventional CCSD, CCSDt, and CCSDT approaches in a strongly correlated regime,
while producing the energies that remain close to those obtained with FCI at all H--H separations
considered in our calculations. Focusing first on the ACP methodologies with up to two-body
clusters, meaning ACCSD(1,3), {\DCSD} = DCSD, {\ACCSDX}, and ACCSD(1,4), the ACCSD(1,3) scheme,
which corresponds to setting $\lambda$ in Eqs.\ \eqref{lambda} and \eqref{accsdt-doubles} at 1,
performs the best, generating a PEC that closely reproduces its exact, FCI, counterpart. This can be
illustrated by the small mean unsigned error (MUE) and mean signed error (MSE) values relative
to FCI characterizing the ACCSD(1,3) PEC, which are 0.675 $\text{mE}_\text{h}$ and
$-0.311$ $\text{mE}_\text{h}$, respectively (see Table \ref{ACP_H6_doubles}). The ACCSD approach
at the other end of the spectrum, \foreign{i.e.}, ACCSD(1,4), which is obtained by using $\lambda = 0$
in Eqs.\ \eqref{lambda} and \eqref{accsdt-doubles}, while being qualitatively correct, does not work 
as well as its ACCSD(1,3) counterpart, increasing the MUE and MSE values relative to FCI to more than
9 $\text{mE}_\text{h}$. As a result, the {\DCSD} = DCSD method, which treats diagrams (3) and (4)
shown Fig.\ \ref{figure1} on equal footing by setting $\lambda$ in Eqs.\ \eqref{lambda} and \eqref{accsdt-doubles}
at $\tfrac{1}{2}$, produces a PEC that is more or less the average of the PECs obtained in the ACCSD(1,3)
and ACCSD(1,4) calculations [see panels (a) and (d) of Fig.\ \ref{figure2} and Table \ref{ACP_H6_doubles}].
This should be contrasted by the {\ACCSDX} computations, in which $\lambda$ in
Eqs.\ \eqref{lambda} and \eqref{accsdt-doubles} is set at $\tfrac{n_\text{o}}{n_\text{o} + n_\text{u}}$
and which result in a PEC similar to that obtained with the ACCSD(1,4) approach, improving the ACCSD(1,4)
energetics only slightly. This behavior of the {\ACCSDX} method can be understood if we realize that the
numbers of occupied and unoccupied orbitals involved in the calculations for the $\text{H}_{6}$/cc-pVTZ system
are 3 and 81, respectively, so that $n_\text{u} = 27 n_\text{o} \gg n_\text{o}$, making the scaling factors
at diagrams (3) and (4) in Eqs.\ \eqref{lambda} and \eqref{accsdt-doubles} close to 0 and 1, respectively,
and the {\ACCSDX} approach similar to ACCSD(1,4).

Based on the above discussion, one might crown the ACCSD(1,3) scheme the best ACP approach examined
in the present study, but this would be misleading. Given the total neglect of $T_{3}$ correlations
in the ACCSD(1,3) (and all other ACCSD) calculations, the excellent performance of the ACCSD(1,3) method
in reproducing the FCI energetics characterizing the metal--insulator transition in the hexagonal
$\text{H}_{6}$/cc-pVTZ system discussed here is fortuitous. As explained in Section \ref{sec2.1}, the ACP approaches
can be very effective in capturing the non-dynamical correlation effects associated with the entanglement
of larger numbers of electrons by taking advantage of the cancellations of certain $(T_{2})^{2}$
diagrams in the CCD/CCSD equations projected on the doubly excited CSFs or determinants in
a strongly correlated regime modeled by the Hubbard and PPP Hamiltonians. However, as pointed out
above, in Section \ref{sec2.1} as well, the diagram cancellations within the CCD/CCSD amplitude equations
that result in the ACP methods of the ACCSD(1,3), ACCSD(1,4), or {\DCSD} = DCSD type do not describe the $T_3$
physics needed to capture much of the remaining dynamical and non-dynamical correlations and obtain a quantitative
description of realistic systems described by \foreign{ab initio} Hamiltonians. It is, therefore,
essential to examine what happens when the various ACCSD approaches considered in this study are
embedded in the CCSDt/CCSDT amplitude equations following the recipe discussed in Section \ref{sec2.2}.

As shown in Tables \ref{ACP_H6_active} and \ref{ACP_H6_fullt} and panels (b), (c), (e), and (f)
in Fig.\ \ref{figure2}, the inclusion of the connected triply excited clusters in the ACP schemes, following the
procedure described by Eqs.\ \eqref{ccsdt-singles}, \eqref{ccsdt-triples}, and \eqref{accsdt-doubles},
has interesting effects on the ACCSD(1,3), {\DCSD} = DCSD, {\ACCSDX}, and ACCSD(1,4) energetics.
The selection of the $(T_{2})^{2}$ diagrams (1) and (3) of Fig.\ \ref{figure1}, which produced
the most accurate PEC for the symmetric dissociation of the $\text{H}_{6}$/cc-pVTZ system among
the various ACCSD methods considered in this work, results in the worst description when the
ACCSDT(1,3), {\DCSDT}, {\ACCSDTX}, and ACCSDT(1,4) PECs and their analogs obtained in the
active-space ACCSDt(1,3), {\DCSDt}, {\ACCSDtX}, and ACCSDt(1,4) calculations are compared with
one another. To make matters worse, the ACCSDt(1,3) and ACCSDT(1,3) approaches, in spite of accounting
for $T_{3}$ correlations, are considerably less accurate than their ACCSD(1,3) counterpart, in which
the connected triple excitations are ignored. This is in sharp contrast with the remaining
ACCSDt and ACCSDT methods considered in this study, which all improve the underlying ACCSD results.
There is, however, a significant difference between the {\ACCSDTX} and ACCSDT(1,4) approaches
and their active-space {\ACCSDtX} and ACCSDt(1,4) counterparts, which are obtained by
including the connected triply excited clusters on top of {\ACCSDX} and ACCSD(1,4),
respectively, and the {\DCSDT} and {\DCSDt} approximations that build upon {\DCSD} = DCSD.
The ACCSD(1,4), ACCSDt(1,4), and ACCSDT(1,4) methods form a systematically improvable
hierarchy, with the MUE values relative to FCI decreasing from 9.349 $\text{mE}_\text{h}$
in the case of ACCSD(1,4) to 2.129 $\text{mE}_\text{h}$, when the ACCSDt(1,4) approximation to
ACCSDT(1,4) is employed, and to 1.304 $\text{mE}_\text{h}$ in the ACCSDT(1,4) case.
The {\ACCSDX}, {\ACCSDtX}, and {\ACCSDTX} approaches behave in a similar manner,
with the corresponding MUE values relative to FCI decreasing from 9.037 $\text{mE}_\text{h}$ to
1.818 $\text{mE}_\text{h}$ and 1.150 $\text{mE}_\text{h}$, respectively, further improving the
ACCSDt(1,4) and ACCSDT(1,4) results, but the analogous DCSD-based sequence, \foreign{i.e.},
{\DCSD} = DCSD, {\DCSDt}, and {\DCSDT}, is no longer as systematic. Indeed, while the
{\DCSDt} and {\DCSDT} calculations that include information about $T_{3}$ clusters are more accurate
than their {\DCSD} counterpart, in which the effects of connected triples are neglected,
the improvements offered by the {\DCSDt} and {\DCSDT} methods compared to {\DCSD} are rather small,
reducing the MUE value relative to FCI characterizing the {\DCSD} calculations, of 4.770 $\text{mE}_\text{h}$,
to 2.585 $\text{mE}_\text{h}$ and 3.486 $\text{mE}_\text{h}$, respectively. Furthermore, unlike
in the ACCSDt(1,4) \foreign{vs.} ACCSDT(1,4) and {\ACCSDtX} \foreign{vs.} {\ACCSDTX} computations,
we do not observe improvements in the results, when the active-space {\DCSDt} approach is replaced
by its {\DCSDT} parent. One should always observe improvements, when going from the CCSDt-level
using small numbers of active orbitals to its CCSDT-type counterpart in larger basis set
calculations, but this is not the case when
the PECs describing the symmetric dissociation of the $\text{H}_{6}$/cc-pVTZ system obtained
with the {\DCSDt} and {\DCSDT} methods are compared with each other. The ACCSDt(1,4)/ACCSDT(1,4)
and, especially, the {\ACCSDtX}/{\ACCSDTX} pairs behave more systematically in this regard.

It is clear from the results presented in Tables \ref{ACP_H6_active} and \ref{ACP_H6_fullt} and
panels (b), (c), (e), and (f) of Fig.\ \ref{figure2} that from all $T_{3}$-corrected CC and ACP
methods investigated in this study, the {\ACCSDtX} and {\ACCSDTX} approaches provide the most accurate
description of the dissociating $\text{H}_{6}$/cc-pVTZ ring that undergoes a transition from the weakly
correlated metallic phase near the equilibrium geometry to the strongly correlated insulator at
larger values of $R_{\text{H--H}}$. They eliminate dramatic failures of
CCSDt and CCSDT at larger H--H separations, they significantly improve the underlying {\ACCSDX} PEC, and they
offer impressive, millihartree-type, accuracies in the entire region of the $R_{\text{H--H}}$ values examined
in this work, recovering 99--100 \% of the FCI correlation energy independent of $R_{\text{H--H}}$.
It is also encouraging to observe that the $n_{\rm o}$- and $n_{\rm u}$-dependent diagram
scaling factors adopted in the {\ACCSDtX} and {\ACCSDTX} schemes improve the {\DCSDt} and {\DCSDT} results
and their ACCSDt(1,3)/ACCSDT(1,3) and ACCSDt(1,4)/ACCSDT(1,4) counterparts. To appreciate the
significance of these findings even more, one should point out that the costs of the {\ACCSDtX}
computations for the $\text{H}_{6}$/cc-pVTZ system are not much greater than those characterizing
the CCSD and ACCSD approximations. Indeed, we needed 2 s per iteration on a single core of the
aforementioned Dell architecture to perform the CCSD or ACCSD calculations, which should be compared to
6 s required by the corresponding {\ACCSDtX} runs. The {\ACCSDTX} method, which offers a full treatment of $T_{3}$
correlations, improving the already excellent {\ACCSDtX} PEC even further, offers promise too.
Obviously, the {\ACCSDTX} approach, using all triply excited cluster amplitudes, is less practical
than its active-space {\ACCSDtX} counterpart, which uses their small fraction, but with less than 100 s
per iteration on a single core of Dell machine used in our computations for the $\text{H}_{6}$/cc-pVTZ
system, it is orders of magnitude less expensive than the exact Hamiltonian diagonalizations, while providing
similar results. Indeed, our FCI calculations for the $\text{H}_{6}$/cc-pVTZ system required more than
7 h per iteration on the same core, even though they exploited the $D_\text{2h}$ symmetry that was not
used in our CC and ACP runs.
Having said all this, the most important finding of our calculations for the symmetric dissociation of
the $\text{H}_{6}$/cc-pVTZ ring is the observation that the active-space {\ACCSDtX} approach provides the
results of the near-FCI quality, capturing virtually all of the relevant dynamical and non-dynamical
correlations, even at larger H--H separations, faithfully reproducing the accurate parent {\ACCSDTX}
data at the small fraction of the computational cost, and reducing the timings of the exact Hamiltonian
diagonalizations from hours per iteration to seconds.

\begin{figure*}
	\centering
	\includegraphics[scale=1]{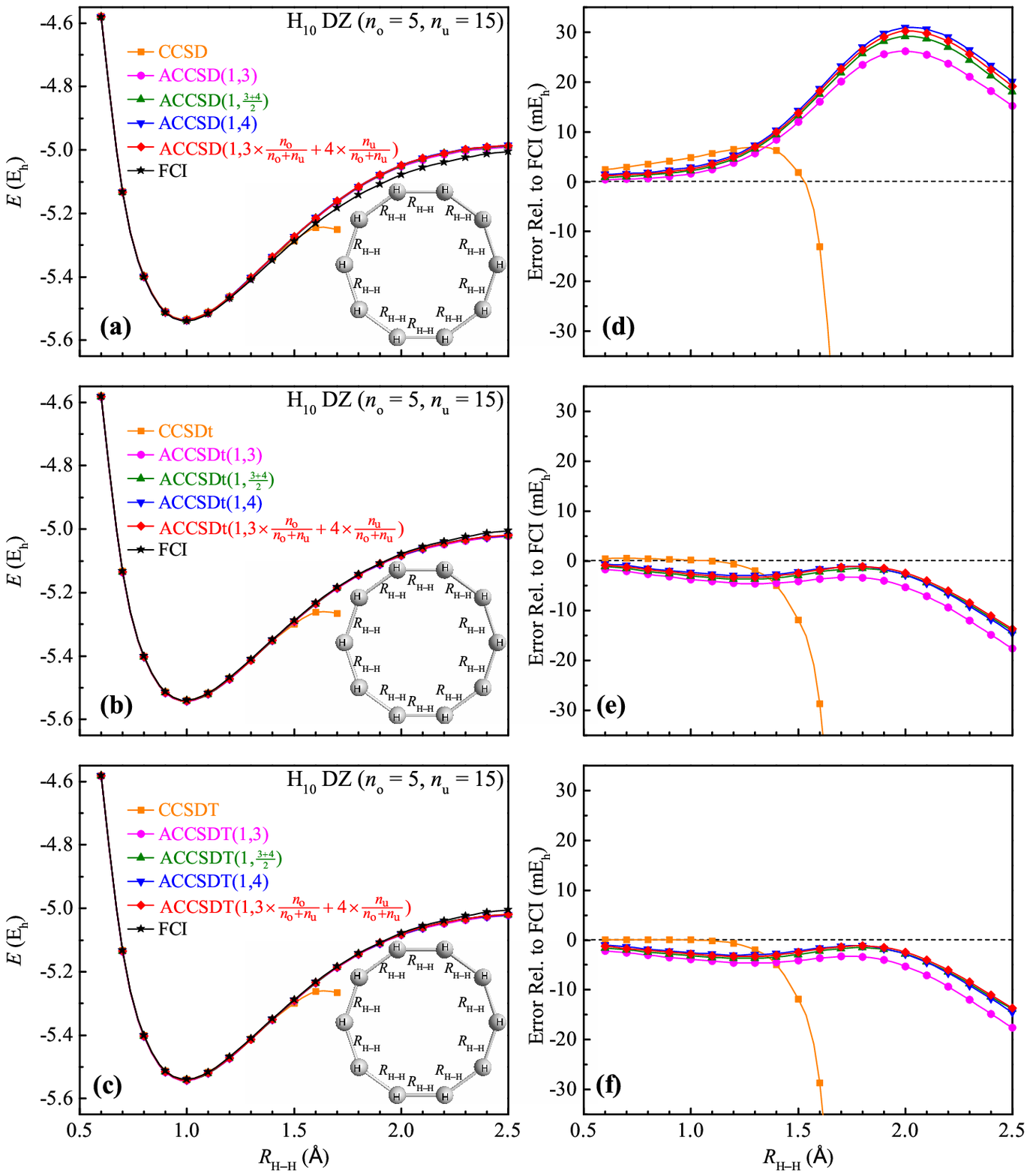}
	\caption{\label{figure3}
		Ground-state PECs [panels (a)--(c)] and errors relative 
		to FCI [panels (d)--(f)] for the symmetric dissociation 
		of the $\text{H}_{10}$ ring resulting from the CCSD and various ACCSD
		[panels (a) and (d)], CCSDt and various ACCSDt 
		[panels (b) and (e)], and CCSDT and various ACCSDT
		[panels (c) and (f)] calculations using the DZ basis set.
                The CCSDt and ACCSDt
		approaches employed a minimum active space consisting of five occupied and five
                lowest-energy unoccupied MOs that correlate with the 1$s$ shells
		of the individual hydrogen atoms. The FCI PEC included in panels 
		(a)--(c) is shown to facilitate comparisons. Note that the CCSD, CCSDt, and CCSDT
                calculations failed to converge in the $R_\text{H--H} \ge 1.8$ {\AA} region.
	}
\end{figure*}

To further investigate the performance of the ACP methods, especially those that incorporate $T_3$ physics,
in a strongly correlated regime coupled with non-trivial dynamical correlations, we studied the
$D_\text{10h}$-symmetric dissociation of the
$\text{H}_{10}$
ring, as described by the DZ basis set.
The ground-state PECs of the $\text{H}_{10}$/DZ system resulting from the CCSD, CCSDt, and CCSDT calculations,
the corresponding ACCSD, ACCSDt, and ACCSDT runs, and the exact FCI computations are shown in Tables
\ref{ACP_H10_doubles}--\ref{ACP_H10_fullt} and Fig.\ \ref{figure3}. Table \ref{ACP_H10_doubles} summarizes
the various CCSD-type calculations, including CCSD, ACCSD(1,3), {\DCSD} = DCSD, {\ACCSDX}, and ACCSD(1,4). The
results of the CCSDt, ACCSDt(1,3), {\DCSDt}, {\ACCSDtX}, and ACCSDt(1,4) computations
are shown in Table \ref{ACP_H10_active}. The CCSDT, ACCSDT(1,3), {\DCSDT}, {\ACCSDTX}, and ACCSDT(1,4)
calculations are summarized in Table \ref{ACP_H10_fullt}.

The symmetric dissociation of the $\text{H}_{10}$ system is more challenging than the previously
discussed $\text{H}_{6}$ ring, since the ground electronic state of $\text{H}_{10}$, when all H--H
distances are simultaneously stretched, involves the entanglement of 10 electrons, which is an
increase in the number of strongly correlated electrons by two thirds.
Because of the use of a smaller basis set, the dimension of the many-electron Hilbert space
characterizing the $D_\text{10h}$-symmetric $\text{H}_{10}$/DZ system is smaller compared to
$\text{H}_{6}$ using the cc-pVTZ basis [the number of the $S_{z} = 0$ determinants of the $A_{g}(D_\text{2h})$
symmetry used in our FCI GAMESS calculations for the $\text{H}_{10}$/DZ system was 60,095,104, as
opposed to 1,139,812,264 used for $\text{H}_{6}$/cc-pVTZ], but the challenge of
strong correlations created by the simultaneous dissociation of all H--H bonds in the
$\text{H}_{10}$ ring is more severe than in the case of its hexagonal $\text{H}_{6}$
cousin. This can be seen when examining the performance of the CCSD, CCSDt, and CCSDT approaches.
As shown in Tables \ref{ACP_H10_doubles}--\ref{ACP_H10_fullt} and Fig.\ \ref{figure3},
the PECs resulting from the CCSD, CCSDt, and CCSDT calculations are not only incorrectly
shaped, exhibiting an unphysical hump at the intermediate stretches of the H--H bonds,
but they cannot be continued beyond $R_{\text{H--H}} \approx 1.7$ {\AA}. The CCSDt and CCSDT methods
reduce the $\sim$4--7 $\text{mE}_\text{h}$ errors relative to FCI obtained with the CCSD approach in the
equilibrium ($R_{\text{H--H}} \approx 1.0$ {\AA}) region to small fractions of a millihartree,
but when $R_{\text{H--H}} > 1.7$ {\AA}, the CCSD, CCSDt, and CCSDT calculations diverge,
even when initiated from the converged cluster amplitudes obtained at the neighboring
geometries from the $R_{\text{H--H}} \leq 1.7$ {\AA} region.
As argued in Refs.\ \ocite{Paldus1984c,Piecuch1990}, using detailed mathematical analysis
of the analytic properties of non-linear CC equations and the relevant numerical evidence,
this is, most likely, caused by the existence of
branch point singularities in the vicinity of the $R_{\text{H--H}} \approx 1.7$ {\AA}
geometry and the CCSD, CCSDt, and CCSDT solutions becoming complex when
$R_{\text{H--H}} > 1.7$ {\AA}. The singularities and divergences of this kind were observed in the
conventional CC calculations for the cyclic polyene models with 14 or more carbon sites 
\cite{Paldus1984c,Paldus1984b,Takahashi1985a,Piecuch1990,Piecuch1991,Paldus1992a,Piecuch1992,Podeszwa2002},
and we believe that our CCSD, CCSDt, and CCSDT computations for the $D_\text{10h}$-symmetric
dissociation of the $\text{H}_{10}$ ring suffer from similar problems once the region of larger
H--H separations is reached.

Tables \ref{ACP_H10_doubles}--\ref{ACP_H10_fullt} and Fig.\ \ref{figure3} demonstrate that all
ACP methods investigated in this work are capable of eliminating the singular behavior observed
in the CCSD, CCSDt, and CCSDT calculations for the dissociating $\text{H}_{10}$ system. As was the
case with the smaller $\text{H}_\text{6}$ ring, the ACCSD(1,3), {\DCSD} = DCSD, {\ACCSDX}, and ACCSD(1,4)
approaches and their ACCSDt and ACCSDT counterparts produce the qualitatively correct PECs that are
in reasonable agreement with FCI, even when the H--H bonds in $\text{H}_{10}$ are significantly stretched.
Although the DZ basis set used in our calculations for the $\text{H}_{10}$ system, in which $n_\text{o} = 5$
and $n_\text{u} = 15$, is much smaller than the cc-pVTZ basis used in the case of $\text{H}_\text{6}$, many
of the accuracy patterns observed in the ACP computations for the $\text{H}_\text{6}$ ring apply to
the symmetric dissociation of its ten-membered analog. For example, among the ACP schemes truncated at
$T_{2}$ examined in this study, the ACCSD(1,3) method is most accurate and the ACCSD(1,4) approach performs worst,
increasing the MUE and MSE values relative to FCI obtained with ACCSD(1,3), of 12.572 $\text{mE}_\text{h}$,
to 15.505  $\text{mE}_\text{h}$ [see Table \ref{ACP_H10_doubles} and panels (a) and (d) in Fig.\ \ref{figure3}].
In analogy to the hexagonal $\text{H}_\text{6}$ system, the {\DCSD} = DCSD and {\ACCSDX} methods,
which average the contributions from diagrams (3) and (4) of Fig.\ \ref{figure1}, produce the PECs
located between the ACCSD(1,3) and ACCSD(1,4) ones. The {\ACCSDX} PEC is closer to that obtained in the
ACCSD(1,4) calculations, since the $\lambda$ factor entering Eqs.\ \eqref{lambda} and \eqref{accsdt-doubles} used
by {\DCSD} is $\tfrac{1}{2}$, whereas the {\ACCSDX} approach proposed in this work adopts the
weighted average defined by $\lambda = \tfrac{n_\text{o}}{n_\text{o} + n_\text{u}}$,
which in the case of the $\text{H}_{10}$/DZ system considered here corresponds to setting $\lambda$
at $\tfrac{1}{4}$, increasing the relative contribution of diagram (4) in
Eqs.\ \eqref{lambda} and \eqref{accsdt-doubles}. Compared to the $\text{H}_\text{6}$/cc-pVTZ
system, the differences among the ACCSD(1,3), {\DCSD}, {\ACCSDX}, and ACCSD(1,4) PECs corresponding
to the symmetric dissociation of the $\text{H}_{10}$ ring are smaller, which is a consequence
of the use of a smaller DZ basis set in the latter case [in the extreme limit of a minimum basis,
where the p--h symmetry is approximately satisfied, the ACCSD(1,3), {\DCSD}, {\ACCSDX}, and ACCSD(1,4)
results would be virtually identical], but the overall accuracy patterns characterizing the ACP methods
with up to two-body clusters, especially the superior performance of the ACCSD(1,3) approximation, hold.

Because of the use of a smaller DZ basis in our calculations for the $\text{H}_{10}$ system, the
similarities in the performance of the ACP methods using different ways of mixing diagrams (3) and (4) in Eqs.\
\eqref{lambda} and \eqref{accsdt-doubles} are also observed at the ACCSDt and ACCSDT
levels, but, as shown in Tables \ref{ACP_H10_active} and \ref{ACP_H10_fullt} and panels (b), (c),
(e), and (f) of Fig.\ \ref{figure3}, the PECs resulting from the {\ACCSDtX} and {\ACCSDTX} calculations
are most accurate when the H--H distances are larger.
All of the ACP approaches with the connected triply excited clusters examined
in this work, including ACCSDT(1,3), {\DCSDT}, {\ACCSDTX}, and ACCSDT(1,4) and their active-space
ACCSDt(1,3), {\DCSDt}, {\ACCSDtX}, and ACCSDt(1,4) counterparts, offer major improvements in the
underlying ACCSD(1,3), {\DCSD} = DCSD, {\ACCSDX}, and ACCSD(1,4) results, reducing the MUE and MSE values
relative to FCI by factors of $\sim$2--4, but, as in the case of the symmetric dissociation of
the $\text{H}_{6}$ ring, the {\ACCSDtX} and {\ACCSDTX} methods perform best overall.
Independent of the H--H
separation, including the $R_{\text{H--H}}$ values as large as 2.5 {\AA} (we recall that the equilibrium
value of $R_{\text{H--H}}$ is about 1.0 {\AA}), they reproduce the FCI correlation energies
to within 1--2 \% or about 3.8--3.9 $\text{mE}_\text{h}$ on average. If we limit ourselves to the
$R_{\text{H--H}} = 0.6\text{--}2.0$ {\AA} region, the errors in the {\ACCSDtX} and {\ACCSDTX} energies
relative to FCI are in the $\sim$1--3 $\text{mE}_\text{h}$ range. Given the dramatic failure and the
singular behavior of the conventional CCSDt and CCSDT approaches and the MUE values characterizing
the {\ACCSDtX} and {\ACCSDTX} calculations being 3--4 times smaller compared to ACCSD(1,3), {\DCSD},
{\ACCSDX}, and ACCSD(1,4), we are encouraged by the performance of the {\ACCSDtX} and {\ACCSDTX} 
schemes developed in this work. It is also encouraging that the unscaled ACCSDt(1,4) and ACCSDT(1,4)
methods, which correspond to setting $\lambda$ in Eqs.\ \eqref{lambda} and \eqref{accsdt-doubles} at 0
and which build upon the old ACCSD(1,4) approximation proposed in Ref.\ \ocite{Piecuch1991} by
incorporating $T_{3}$ correlations in it, are practically as accurate. In particular, they improve
the results obtained with the {\DCSDt} and {\DCSDT} approaches, which extend the {\DCSD} = DCSD scheme
to the CCSDt/CCSDT-type treatments of the connected triples.

Last but not least, our calculations
for the symmetric dissociation of the $\text{H}_{10}$ ring system summarized in Tables
\ref{ACP_H10_active} and \ref{ACP_H10_fullt} and panels (b), (c), (e), and (f) confirm the previous
observation that the active-space ACCSDt methods faithfully reproduce the corresponding ACCSDT energetics,
even at larger H--H separations, at the small fraction of the computational cost.
The differences between the ACCSDt(1,3), {\DCSDt}, {\ACCSDtX}, and ACCSDt(1,4) results and their
respective ACCSDT(1,3),
\linebreak
{\DCSDT}, {\ACCSDTX}, and ACCSDT(1,4) parents obtained for the
$\text{H}_{10}$ system are smaller than in the case of the previously discussed $\text{H}_{6}$ ring,
since the DZ basis set used in our calculations for $\text{H}_{10}$ is much smaller than the cc-pVTZ
basis employed in the $\text{H}_{6}$ case and, as a consequence, the number of active unoccupied orbitals
used in the ACCSDt calculations for the $\text{H}_{10}$/DZ system is a significant fraction
(one third) of all unoccupied MOs, but we still see substantial reduction in the computational
effort offered by the active-space treatment of triples compared to the corresponding ACCSDT levels,
from 2 s per iteration on a single core of the aforementioned Dell node in the case of ACCSDT
to 0.7 s in the ACCSDt case.

\subsection{Symmetric Dissociation of the  \mbox{\boldmath{${\text{H}}_{50}$}} Linear Chain}
\label{sec3.2}

Having analyzed the performance of the various ACP approaches, especially those incorporating the connected
three-body clusters, in calculations for the symmetric dissociations of the six- and ten-membered hydrogen
rings, we proceed to the examination of our final example, which is the $\text{H}_{50}$ linear chain.
As in the case of the $\text{H}_{6}$ and $\text{H}_{10}$ systems discussed in Section \ref{sec3.1},
we model the transition from the weakly correlated metallic regime in the equilibrium region, where
the distances between the adjacent hydrogen atoms are around 1.8 bohr, to the strongly correlated insulator
phase by simultaneously stretching all H--H bonds. Putting aside the periodic nature of the $\text{H}_{6}$
and $\text{H}_{10}$ ring structures, the main difference between the $\text{H}_{6}$ and $\text{H}_{10}$ clusters
and the $\text{H}_{50}$ linear chain is a much larger number of the strongly correlated electrons in the latter 
system, which results in an enormous many-electron Hilbert space, spanned by $\sim 10^{27}$ singlet CSFs
when the minimum basis set is employed and even larger numbers of CSFs when larger basis sets are considered. As
already alluded to above, the exact Hamiltonian diagonalizations for the many-electron Hilbert spaces of such
dimensionalities are outside reach of the best FCI algorithms
and most powerful computer architectures available today, so in
evaluating the performance of the ACP approaches considered in this work in calculations for the symmetric
dissociation of the $\text{H}_{50}$ linear chain, we relied on the nearly exact PEC obtained in the
LDMRG(500) computations using the STO-6G minimum basis set reported in Ref.\ \ocite{Hachmann2006}.
Because of the use of the minimum basis, for which $n_{\rm o} = n_{\rm u} = N_{\rm o} = N_{\rm u} = 25$, our
ACCSDt and CCSDt calculations for the $\text{H}_{50}$ system are equivalent to their ACCSDT and CCSDT counterparts
and the PECs obtained with the {\ACCSDTX} = {\DCSDT}, ACCSDT(1,3), and ACCSDT(1,4) approaches
are nearly identical, so in the discussion below we concentrate on the {\DCSDT} results and the results
generated with CCSD, CCSDT, and {\DCSD} = DCSD. The DCSD calculations for the symmetric dissociation
of the $\text{H}_{50}$ linear chain, as described by the STO-6G basis set, were also reported in
Ref.\ \ocite{Kats2013a}, but the authors of Ref.\ \ocite{Kats2013a} did not consider $T_{3}$ correlations
in their work. Our computations for the $\text{H}_{50}$/STO-6G system discussed in this section allow us
to investigate if the inclusion of the connected triply excited clusters via the {\DCSDT} scheme
implemented in this study can improve the DCSD PEC presented in Ref.\ \ocite{Kats2013a}.

\begin{figure*}
	\centering
	\includegraphics[scale=0.9]{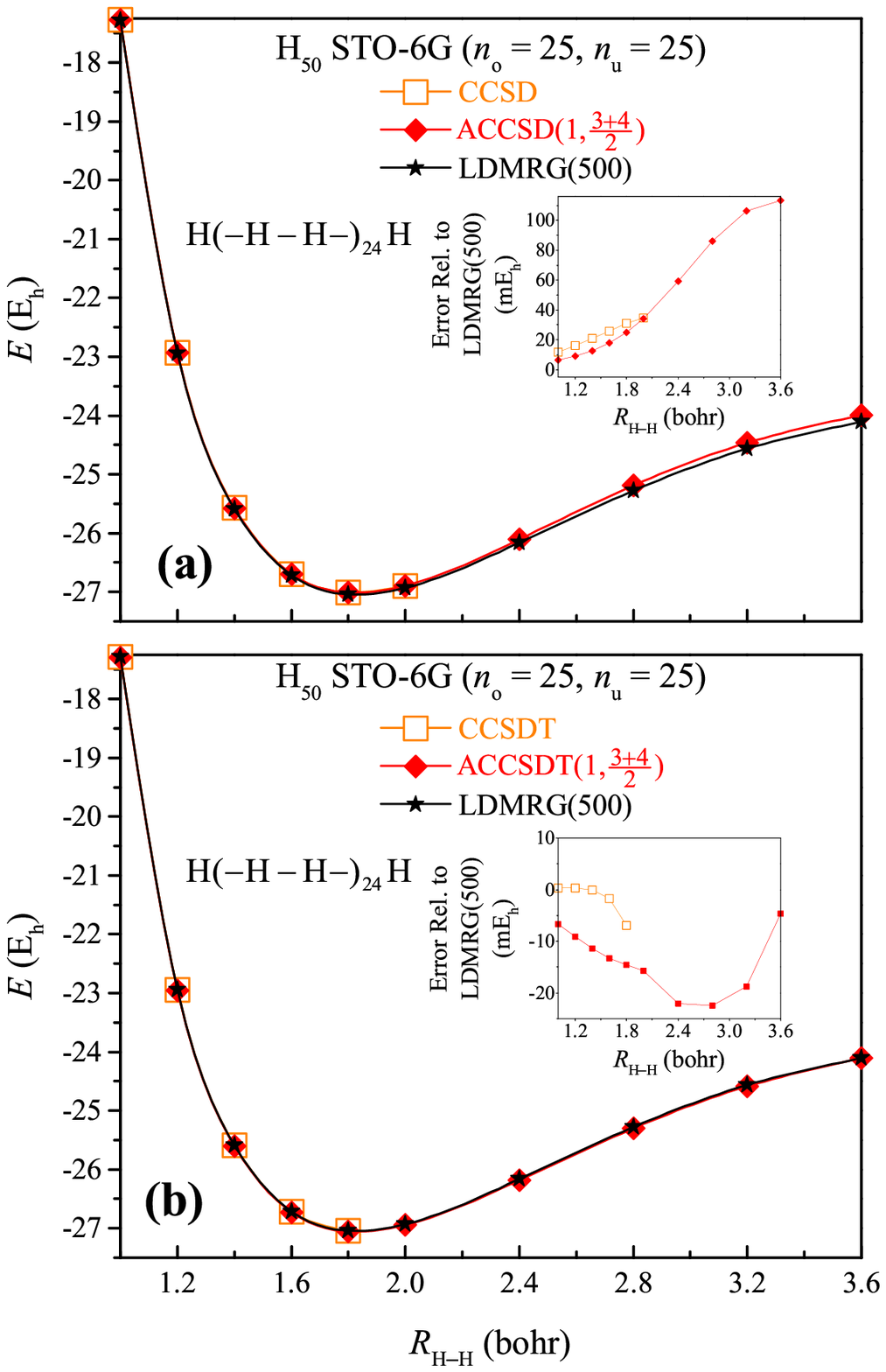}
	\caption{\label{figure4}
		Ground-state PECs for the symmetric dissociation of
		the $\text{H}_{50}$ linear chain obtained in the (a) CCSD and 
		{\DCSD} and (b) CCSDT and {\DCSDT} calculations using the STO-6G basis set.
                In this case $n_{\rm o} = n_{\rm u} = N_{\rm o} = N_{\rm u}$, so that
                {\DCSD} = DCSD is equivalent to {\ACCSDX} and
                {\DCSDT} is equivalent to {\DCSDt}, {\ACCSDtX}, and {\ACCSDTX}.
		The LDMRG(500) PEC included for
                comparison purposes is based on the data reported in Ref.\ \ocite{Hachmann2006}.
		The insets show the errors relative to LDMRG(500).
                Note that the CCSD calculations failed to converge in the $R_\text{H--H} > 2.0$ bohr region.
                We could not converge the CCSDT equations beyond $R_\text{H--H} = 1.8$ bohr.
	}
\end{figure*}

The information about the ground-state PECs corresponding to the symmetric dissociation of the
$\text{H}_{50}$ linear chain, as described by the STO-6G basis, which resulted from our
CCSD, CCSDT, {\DCSD} = DCSD, and {\DCSDT} calculations,
along with the reference LDMRG(500) energetics taken from Ref.\ \ocite{Hachmann2006}, is
summarized in Table \ref{ACP_H50} and Fig.\ \ref{figure4}. As one might anticipate,
given the large number of strongly correlated electrons, the $\text{H}_{50}$ system
is much more challenging than the previously discussed $\text{H}_{6}$ and $\text{H}_{10}$ rings.
This becomes apparent when we look at the results of the CCSD and CCSDT calculations,
which fail to converge already near the equilibrium region, most likely due to the
emergence of branch point singularities and the real solutions of the corresponding
amplitude equations becoming complex
as all H--H bonds of the $\text{H}_{50}$ linear chain are simultaneously stretched.
As pointed out in Section \ref{sec3.1}, the singularities and divergences of this kind
were observed in the studies of cyclic polyene models reported in Refs.\
\ocite{Paldus1984c,Paldus1984b,Takahashi1985a,Piecuch1990,Piecuch1991,Paldus1992a,Piecuch1992,Podeszwa2002}.
The fact that we could not converge the CCSD and CCSDT amplitudes for the $\text{H}_{50}$ system
already in the equilibrium region is consistent with the observations made in Refs.\
\ocite{Paldus1984c,Piecuch1990,Piecuch1991,Paldus1992a,Podeszwa2002}
that the branch point singularities plaguing conventional CC calculations move toward
the regions of weaker correlations as the system size increases. The CCSDT method
offers significant improvements in the CCSD energies at compressed H--H distances, but
this is not helpful, since neither CCSD nor CCSDT solutions can be continued
beyond the equilibrium region.

In view of the above failures of the conventional CCSD and CCSDT approaches, it is
encouraging to see that our {\DCSDT} scheme provides a non-singular and reliable description
of the $\text{H}_{50}$ linear chain in the entire range of H--H separations considered
in this work, including the H--H bond lengths as large as twice the equilibrium value of
$R_{\text{H--H}}$, reducing the MUE relative to the LDMRG(500) reference data that characterizes
the already well-behaved {\DCSD} = DCSD method, of more than 47 $\text{mE}_\text{h}$,
to less than 14 $\text{mE}_\text{h}$ (see Table \ref{ACP_H50} and Fig.\ \ref{figure4}).
The improvements in the results obtained with the lower-rank {\DCSD} approach, which
neglects $T_{3}$ contributions, offered by the {\DCSDT} method, which includes them,
are especially significant when $R_{\text{H--H}} \geq 2.0$ bohr, \foreign{i.e.},
when the CCSDT calculations stop converging. While it is possible that the PEC
representing the symmetric dissociation of the $\text{H}_{50}$ linear chain obtained
with {\DCSDT} starts behaving erratically when $R_{\text{H--H}} \gg 3.6$ bohr, the error
reductions compared to {\DCSD} in the strongly correlated $R_{\text{H--H}} = 2.0\text{--}3.6$ bohr
region, from about 34--113 $\text{mE}_\text{h}$ to 5--22 $\text{mE}_\text{h}$, are quite
remarkable. The errors relative to the nearly exact LDMRG(500) data obtained in the
{\DCSDT} calculations for the $\text{H}_{50}$/STO-6G system are larger, in absolute
value, than the errors relative to FCI characterizing the {\DCSDT} and other
ACCSDT computations for the $\text{H}_{6}$ and $\text{H}_{10}$ rings,
but we have to keep in mind that $\text{H}_{50}$ is much larger
than $\text{H}_{6}$ and $\text{H}_{10}$, so the errors relative to exact or nearly
exact computations are expected to be larger too. What is more important is the observation that
the {\DCSDT} calculations for the dissociating $\text{H}_{50}$/STO-6G linear chain reproduce
the LDMRG(500) correlation energies, which we use as a reference in this study, to within 1--2 \%.
This is a noticeable improvement over the underlying {\DCSD} approach, which gives
2--5 \% differences with the correlation energies obtained in the LDMRG(500) calculations
reported in Ref.\ \ocite{Hachmann2006}. It is encouraging to observe that the {\DCSDT}
approach and most of the other methods in this category examined in this work, including
{\ACCSDTX} and ACCSDT(1,4) and their active-space variants, are capable of reproducing
the exact or nearly exact correlation energies in the strongly correlated systems of the
type of the hydrogen clusters discussed here and in Section \ref{sec3.1} to within 1--2 \%,
independent of the system size and size of the basis employed in the calculations, substantially improving
the corresponding ACCSD computations.
The {\DCSDT} approach is more expensive than its lower-rank {\DCSD} = DCSD counterpart,
increasing 0.8 s per iteration on a single core of the Dell machine used in the present study to 482 s,
when the $\text{H}_{50}$/STO-6G system is examined. We have to realize, however, that
a few minutes per iteration on a single core for a system containing 50 strongly correlated electrons
is an insignificant effort, given the observed improvements in the description of the corresponding PEC
over CCSD, CCSDT, and {\DCSD} offered by {\DCSDT} in the region of stretched H--H distances,
not to mention the fact that FCI calculations for $\text{H}_{50}$ are not feasible.

\section{Conclusions}
\label{sec4}

It is well established that the conventional single-reference CCSD, CCSDT, \foreign{etc.}\ hierarchy, in which
the cluster operator $T$ is truncated at a given many-body rank, displays an erratic behavior when
the number of strongly correlated electrons becomes larger than in typical multi-reference situations
in chemistry involving single or double bond breaking. At the same time, traditional multi-reference
approaches may become inapplicable due to the prohibitively large dimensionalities of the underlying
multi-configurational model spaces when the numbers of active electrons and orbitals become larger.
The ACP theories, in which one uses subsets of non-linear $(T_{2})^{2}$ diagrams in the CCD or CCSD amplitude
equations, are capable of capturing strong correlations involving the entanglement of larger numbers of
electrons, but they neglect the physics associated with the connected triply excited clusters needed for
a more quantitative description. Furthermore, the specific combinations of $(T_{2})^{2}$ diagrams that
work well in the strongly correlated regime of the minimum-basis-set model systems used in the past
to rationalize the ACP methods may not necessarily be optimum when larger basis sets and more substantial
dynamical correlations are involved. The objective of this study has been to examine both topics
and suggest and test solutions that might help develop better and more quantitative ACP models in the future.

In this work, we dealt with the issue of the missing $T_{3}$ physics by combining the selected ACP schemes,
including the previously formulated ACCSD(1,3), ACCSD(1,4), and {\DCSD} = DCSD approaches and the
novel {\ACCSDX} methodology, in which diagrams (3) and (4) of Fig.\ \ref{figure1} are scaled by
factors depending on the numbers of occupied ($n_{\rm o}$) and unoccupied ($n_{\rm u}$)
orbitals, with the idea of capturing the leading triply excited cluster amplitudes using active orbitals,
which was previously exploited in the CCSDt and similar methods.
The resulting ACCSDt(1,3), ACCSDt(1,4), {\DCSDt}, and {\ACCSDtX}
methods, which are obtained by considering the appropriate subsets of $(T_{2})^{2}$ diagrams in the
amplitude equations projected on the doubly excited determinants within the CCSDt system, their
more expensive ACCSDT(1,3), ACCSDT(1,4), {\DCSDT}, and {\ACCSDTX} parents, in which $T_{3}$ component
is treated fully, obtained from CCSDT, and the underlying ACCSD(1,3), ACCSD(1,4), {\DCSD}, and {\ACCSDX}
approximations were implemented and tested using the symmetric dissociations of the
$\text{H}_6$ and $\text{H}_{10}$ rings, as described by the cc-pVTZ and DZ basis sets, and
the $\text{H}_{50}$ linear chain treated with STO-6G. We used these
three systems as our numerical examples, since the linear chains and rings of the
equally spaced hydrogen atoms can be used to model metal--insulator transitions in which
the weakly correlated metallic phase at compressed geometries becomes the strong correlated insulator
at larger H--H separations.
The $T_{3}$-corrected {\ACCSDtX} and {\ACCSDTX} schemes and their lower-rank {\ACCSDX} counterpart,
in which the connected triply excited clusters are neglected,  were used to investigate if
the $n_{\rm o}$- and $n_{\rm u}$-dependent scaling factors at diagrams (3) and (4) can help
to improve accuracies when basis sets are larger than a minimum one. We gauged the performance
of the various ACP schemes examined in this work by comparing the resulting PECs for the
$\text{H}_6$/cc-pVTZ, $\text{H}_{10}$/DZ, and $\text{H}_{50}$/STO-6G systems
with their exact, FCI ($\text{H}_6$/cc-pVTZ and $\text{H}_{10}$/DZ) or nearly exact, DMRG
($\text{H}_{50}$/STO-6G) counterparts. The FCI PECs for the symmetric dissociations of the
$\text{H}_6$/cc-pVTZ and $\text{H}_{10}$/DZ ring systems were generated in this work using GAMESS.
The reference PEC for the symmetric dissociation of the $\text{H}_{50}$ linear chain, as described
by the STO-6G basis, was taken from Ref.\ \ocite{Hachmann2006}, where the authors used
the LDMRG(500) approach. We also included the conventional CCSD, CCSDt, and CCSDT methods
in our calculations.

We demonstrated that all ACP approaches examined in this study, without and with the connected triples,
provide qualitatively correct PECs, eliminating the erratic and, in the case of $\text{H}_{10}$ and
$\text{H}_{50}$, singular behavior of CCSD, CCSDt, and CCSDT in the strongly correlated regime.
Of all ACP methods including $T_{3}$ correlations, the {\ACCSDtX} approach and its {\ACCSDTX} parent,
which in the minimum-basis-set case become equivalent to {\DCSDt} and {\DCSDT}, respectively, turned out
to be overall most accurate,
improving the corresponding {\ACCSDX} [in the case of {\DCSDt} and {\DCSDT},
{\DCSD} = DCSD] PECs, especially in the regions of larger H--H separations, where correlations
become stronger. In general, it was promising to find out that the {\ACCSDTX}, ACCSDT(1,4), and {\DCSDT}
methods and their more economical active-space {\ACCSDtX}, ACCSDt(1,4), and {\DCSDt} counterparts
reproduce the exact or nearly exact correlation energies in the hydrogen clusters,
which model transitions from the weakly correlated metallic regime to the strongly correlated insulator
phase, to within 1--2 \%, reducing errors in the underlying ACCSD computations in a substantial manner.
It was also encouraging to observe that all ACCSDt schemes, which use active orbitals that
reflect on the nature of strong correlations of interest to identify the dominant $T_{3}$ contributions,
faithfully reproduce the results of the corresponding ACCSDT calculations, in which the connected three-body
clusters are treated fully, at the small fraction of the computational effort. While the 
$n_{\rm o}$- and $n_{\rm u}$-dependent scaling factors at diagrams (3) and (4) did not help
the ACCSD calculations, they seem to be useful in improving the accuracies of the ACCSDt and ACCSDT
computations when larger basis sets are employed. Having said this, there may be other, better ways of
improving accuracies of $T_{3}$-corrected ACP approaches in calculations for strongly correlated systems
involving the entanglement of larger numbers of electrons and using larger basis sets, and we will
search for such ways in the future. 

Among the other issues we would like to pursue is search for diagram cancellations within the externally corrected
CCSDt and CCSDT amplitude equations projected on triples that would be consistent with the $(T_{2})^{2}$ diagram
selections used by the ACP methods in the equations projected on doubles. While some diagram selections within the
CCSDT equations were considered in Refs.\ \ocite{Bartlett2006,Musial2007,Kats2019,Kats2021,Rishi2019},
it remains unclear if the resulting or similar
methods can handle strongly correlated systems involving the entanglement of larger numbers of electrons.
As pointed out in Section \ref{sec2}, model systems that were used in the past to justify the $(T_{2})^{2}$
diagram cancellations defining the ACP schemes by extracting $T_{4}$ contributions from the wave functions
obtained by projecting out the singlet components of the UHF determinants breaking the $S^{2}$ symmetry, but
not $S_{z}$, do not offer any guidance how to handle $T_{3}$ clusters, since the $T_{n}$
components with odd values of $n \geq 3$ are absent in the resulting PUHF states \cite{Paldus1984a,Piecuch1996a}.
In order to see the emergence of $T_{3}$ contributions through spin-symmetry breaking and restoration, one
has to break the $S^{2}$ as well as $S_{z}$ symmetries, as in Ref.\ \ocite{Henderson2017}. Thus, it would be
worth examining if the considerations reported in Ref.\ \ocite{Henderson2017} allow one to come up with
diagram cancellations or combinations within the CCSDt/CCSDT amplitude equations that could describe
strong correlations, as understood throughout the present article, while accounting for the connected
triply excited clusters at the same time.


\section*{Funding}

This work has been supported by the Chemical Sciences, Geosciences
and Biosciences Division, Office of Basic Energy Sciences, Office
of Science, U.S. Department of Energy (Grant No. DE-FG02-01ER15228).

\section*{ORCID}

\noindent
Ilias Magoulas \verb"https://orcid.org/0000-0003-3252-9112" \\
Jun Shen \verb"https://orcid.org/0000-0003-1838-3719" \\
Piotr Piecuch \verb"https://orcid.org/0000-0002-7207-1815"


%

\pagebreak



\squeezetable
\begin{table*}[h!]
\caption{\label{ACP_H6_doubles}
A comparison of the energies obtained with the CCSD and various
ACCSD approaches with the exact FCI 
data for the symmetric dissociation of the $\text{H}_{6}$
ring, as described by the cc-pVTZ basis set,
at selected bond distances between neighboring H atoms 
$R_\text{H--H}$ 
(in \AA).\protect\footnotemark[1]}
\begin{ruledtabular}
\begin{tabular*}{\textwidth}{@{\extracolsep{\fill}} l d d d d d d}
		\multirow{2}{*}{$R_\text{H--H}$} & 
		\multicolumn{1}{c}{\multirow{2}{*}{CCSD}} &
		\multicolumn{4}{c}{ACCSD} &
		\multicolumn{1}{c}{\multirow{2}{*}{FCI}} \\\cline{3-6}
		& &
		\multicolumn{1}{c}{(1,3)} &
		\multicolumn{1}{c}{$(1,\tfrac{3+4}{2})$\protect\footnotemark[2]} & 
		\multicolumn{1}{c}{$(1, 3 \times \tfrac{n_\text{o}}{n_\text{o} + 
				n_\text{u}} + 4 \times \tfrac{n_\text{u}}{n_\text{o} + 
				n_\text{u}})$} &
		\multicolumn{1}{c}{(1,4)} \\[1mm]
\hline
		0.6 &    3.422 & -0.242 & 1.557 &  3.144 &  3.262 & -2.858958 \\
		0.7 &    3.593 & -0.320 & 1.618 &  3.321 &  3.448 & -3.176147 \\
		0.8 &    3.823 & -0.417 & 1.695 &  3.541 &  3.679 & -3.331124 \\
		0.9 &    4.103 & -0.534 & 1.788 &  3.809 &  3.959 & -3.396176 \\
		1.0 &    4.424 & -0.665 & 1.906 &  4.132 &  4.297 & -3.410069 \\
		1.1 &    4.774 & -0.805 & 2.059 &  4.522 &  4.705 & -3.394673 \\
		1.2 &    5.139 & -0.944 & 2.263 &  5.003 &  5.206 & -3.362943 \\
		1.3 &    5.489 & -1.068 & 2.543 &  5.608 &  5.833 & -3.322850 \\
		1.4 &    5.758 & -1.156 & 2.934 &  6.381 &  6.634 & -3.279466 \\
		1.5 &    5.802 & -1.175 & 3.479 &  7.377 &  7.662 & -3.236119 \\
		1.6 &    5.309 & -1.088 & 4.219 &  8.638 &  8.960 & -3.195040 \\
		1.7 &    3.647 & -0.865 & 5.165 & 10.165 & 10.529 & -3.157716 \\
		1.8 &   -0.380 & -0.508 & 6.266 & 11.872 & 12.280 & -3.125051 \\
		1.9 &   -8.826 & -0.071 & 7.388 & 13.563 & 14.012 & -3.097433 \\
		2.0 &  -24.720 &  0.350 & 8.337 & 14.963 & 15.445 & -3.074787 \\
		2.1 &  -51.331 &  0.660 & 8.931 & 15.813 & 16.314 & -3.056686 \\
		2.2 &  -89.837 &  0.803 & 9.073 & 15.975 & 16.478 & -3.042507 \\
		2.3 & -137.197 &  0.782 & 8.776 & 15.463 & 15.951 & -3.031571 \\
		2.4 & -187.824 &  0.635 & 8.135 & 14.417 & 14.876 & -3.023237 \\
		2.5 & -237.074 &  0.416 & 7.277 & 13.027 & 13.447 & -3.016948 \\
\hline
		MUE\protect\footnotemark[3] &   39.624 &  0.675 & 
		4.770 &  
		9.037 &  9.349 & \multicolumn{1}{c}{---} \\
		MSE\protect\footnotemark[4] &  -34.095 & -0.311 & 
		4.770 &  
		9.037 &  9.349 & \multicolumn{1}{c}{---} \\
\end{tabular*}
\end{ruledtabular}
\footnotetext[1]{\setlength{\baselineskip}{1em}
		The FCI energies are total energies in hartree, whereas all of the 
		remaining energies are errors relative to FCI in millihartree.}
\footnotetext[2]{\setlength{\baselineskip}{1em}
                Equivalent to the DCSD approach of Ref.\ \ocite{Kats2013a}.}
\footnotetext[3]{\setlength{\baselineskip}{1em}
		Mean unsigned error.}
\footnotetext[4]{\setlength{\baselineskip}{1em}
		Mean signed error.}
\end{table*}

\begin{table*}[h!]
\caption{\label{ACP_H6_active}
A comparison of the energies obtained with the CCSDt and various
ACCSDt approaches with the exact FCI
data for the symmetric dissociation of the $\text{H}_{6}$
ring, as described by the cc-pVTZ basis set,
at selected bond distances between neighboring H atoms
$R_\text{H--H}$
(in \AA).\protect\footnotemark[1]}
\begin{ruledtabular}
\begin{tabular*}{\textwidth}{@{\extracolsep{\fill}} l d d d d d d}
		\multirow{2}{*}{$R_\text{H--H}$} & 
		\multicolumn{1}{c}{\multirow{2}{*}{CCSDt}} &
		\multicolumn{4}{c}{ACCSDt} &
		\multicolumn{1}{c}{\multirow{2}{*}{FCI}} \\\cline{3-6}
		& &
		\multicolumn{1}{c}{(1,3)} &
		\multicolumn{1}{c}{$(1,\tfrac{3+4}{2})$} & 
		\multicolumn{1}{c}{$(1, 3 \times \tfrac{n_\text{o}}{n_\text{o} + 
				n_\text{u}} + 4 \times \tfrac{n_\text{u}}{n_\text{o} + 
				n_\text{u}})$} &
		\multicolumn{1}{c}{(1,4)} \\ [1mm]
\hline
		0.6 &    2.548 &  -1.234 &  0.615 & 2.243 & 2.365 & -2.858958 \\
		0.7 &    2.419 &  -1.660 &  0.345 & 2.103 & 2.235 & -3.176147 \\
		0.8 &    2.235 &  -2.247 & -0.043 & 1.880 & 2.023 & -3.331124 \\
		0.9 &    2.018 &  -2.962 & -0.515 & 1.608 & 1.766 & -3.396176 \\
		1.0 &    1.804 &  -3.757 & -1.021 & 1.339 & 1.514 & -3.410069 \\
		1.1 &    1.609 &  -4.592 & -1.519 & 1.114 & 1.309 & -3.394673 \\
		1.2 &    1.414 &  -5.454 & -1.992 & 0.953 & 1.170 & -3.362943 \\
		1.3 &    1.111 &  -6.428 & -2.504 & 0.809 & 1.053 & -3.322850 \\
		1.4 &    0.669 &  -7.437 & -2.974 & 0.766 & 1.040 & -3.279466 \\
		1.5 &   -0.074 &  -8.459 & -3.375 & 0.855 & 1.163 & -3.236119 \\
		1.6 &   -1.438 &  -9.451 & -3.673 & 1.104 & 1.451 & -3.195040 \\
		1.7 &   -4.014 & -10.345 & -3.837 & 1.516 & 1.904 & -3.157716 \\
		1.8 &   -8.888 & -11.062 & -3.852 & 2.051 & 2.478 & -3.125051 \\
		1.9 &  -17.983 & -11.538 & -3.744 & 2.611 & 3.069 & -3.097433 \\
		2.0 &  -34.481 & -11.755 & -3.576 & 3.059 & 3.535 & -3.074787 \\
		2.1 &  -62.336 & -11.747 & -3.432 & 3.266 & 3.744 & -3.056686 \\
		2.2 & -102.675 & -11.583 & -3.387 & 3.160 & 3.624 & -3.042507 \\
		2.3 & -150.692 & -11.337 & -3.483 & 2.735 & 3.172 & -3.031571 \\
		2.4 & -199.899 & -11.068 & -3.726 & 2.041 & 2.442 & -3.023237 \\
		2.5 & -246.321 & -10.805 & -4.094 & 1.154 & 1.516 & -3.016948 \\
\hline
		MUE\protect\footnotemark[2] &   42.231 &   7.746 &  
		2.585 & 
		1.818 & 2.129 & \multicolumn{1}{c}{---} \\
		MSE\protect\footnotemark[3] &  -40.649 &  -7.746 & 
		-2.489 & 
		1.818 & 2.129 & \multicolumn{1}{c}{---} \\
\end{tabular*}
\end{ruledtabular}
	\footnotetext[1]{\setlength{\baselineskip}{1em}
		The FCI energies are total energies in hartree, whereas all of the 
		remaining energies are errors relative to FCI in millihartree. The
		CCSDt and ACCSDt approaches employed three active occupied and three 
		active unoccupied MOs corresponding to the 1$s$ shells of the 
		individual H atoms.}
	\footnotetext[2]{\setlength{\baselineskip}{1em}
		Mean unsigned error.}
	\footnotetext[3]{\setlength{\baselineskip}{1em}
		Mean signed error.}
\end{table*}

\begin{table*}[h!]
\caption{\label{ACP_H6_fullt}
A comparison of the energies obtained with the CCSDT and various
ACCSDT approaches with the exact FCI
data for the symmetric dissociation of the $\text{H}_{6}$
ring, as described by the cc-pVTZ basis set,
at selected bond distances between neighboring H atoms
$R_\text{H--H}$
(in \AA).\protect\footnotemark[1]}
\begin{ruledtabular}
\begin{tabular*}{\textwidth}{@{\extracolsep{\fill}} l d d d d d d}
		\multirow{2}{*}{$R_\text{H--H}$} & 
		\multicolumn{1}{c}{\multirow{2}{*}{CCSDT}} &
		\multicolumn{4}{c}{ACCSDT} &
		\multicolumn{1}{c}{\multirow{2}{*}{FCI}} \\\cline{3-6}
		& &
		\multicolumn{1}{c}{(1,3)} &
		\multicolumn{1}{c}{$(1,\tfrac{3+4}{2})$} & 
		\multicolumn{1}{c}{$(1, 3 \times \tfrac{n_\text{o}}{n_\text{o} + 
				n_\text{u}} + 4 \times \tfrac{n_\text{u}}{n_\text{o} + 
				n_\text{u}})$} &
		\multicolumn{1}{c}{(1,4)} \\ [1mm]
\hline
		0.6 &    0.079 &  -3.952 & -1.996 & -0.279 & -0.151 & -2.858958 \\
		0.7 &    0.095 &  -4.233 & -2.123 & -0.277 & -0.139 & -3.176147 \\
		0.8 &    0.116 &  -4.609 & -2.304 & -0.295 & -0.146 & -3.331124 \\
		0.9 &    0.140 &  -5.074 & -2.530 & -0.325 & -0.162 & -3.396176 \\
		1.0 &    0.160 &  -5.621 & -2.792 & -0.356 & -0.175 & -3.410069 \\
		1.1 &    0.166 &  -6.241 & -3.079 & -0.374 & -0.174 & -3.394673 \\
		1.2 &    0.137 &  -6.932 & -3.381 & -0.365 & -0.143 & -3.362943 \\
		1.3 &    0.031 &  -7.692 & -3.686 & -0.307 & -0.059 & -3.322850 \\
		1.4 &   -0.235 &  -8.512 & -3.973 & -0.172 &  0.106 & -3.279466 \\
		1.5 &   -0.823 &  -9.369 & -4.213 &  0.072 &  0.385 & -3.236119 \\
		1.6 &   -2.046 & -10.216 & -4.371 &  0.457 &  0.808 & -3.195040 \\
		1.7 &   -4.493 & -10.982 & -4.411 &  0.989 &  1.380 & -3.157716 \\
		1.8 &   -9.245 & -11.585 & -4.316 &  1.631 &  2.061 & -3.125051 \\
		1.9 &  -18.225 & -11.959 & -4.109 &  2.286 &  2.747 & -3.097433 \\
		2.0 &  -34.617 & -12.084 & -3.855 &  2.817 &  3.296 & -3.074787 \\
		2.1 &  -62.392 & -11.995 & -3.635 &  3.095 &  3.576 & -3.056686 \\
		2.2 & -102.685 & -11.760 & -3.526 &  3.048 &  3.514 & -3.042507 \\
		2.3 & -150.676 & -11.452 & -3.567 &  2.673 &  3.111 & -3.031571 \\
		2.4 & -199.857 & -11.129 & -3.764 &  2.019 &  2.421 & -3.023237 \\
		2.5 & -246.248 & -10.820 & -4.092 &  1.165 &  1.528 & -3.016948 \\
\hline
		MUE\protect\footnotemark[2] &   41.623 &   8.811 &  
		3.486 &  
		1.150 &  1.304 & \multicolumn{1}{c}{---} \\
		MSE\protect\footnotemark[3] &  -41.531 &  -8.811 & 
		-3.486 &  
		0.875 &  1.189 & \multicolumn{1}{c}{---} \\
\end{tabular*}
\end{ruledtabular}
	\footnotetext[1]{\setlength{\baselineskip}{1em}
		The FCI energies are total energies in hartree, whereas all of the 
		remaining energies are errors relative to FCI in millihartree.}
	\footnotetext[2]{\setlength{\baselineskip}{1em}
		Mean unsigned error.}
	\footnotetext[3]{\setlength{\baselineskip}{1em}
		Mean signed error.}
\end{table*}

\begin{table*}[h!]
\caption{\label{ACP_H10_doubles}
A comparison of the energies obtained with the CCSD and various
ACCSD approaches with the exact FCI
data for the symmetric dissociation of the $\text{H}_{10}$
ring, as described by the DZ basis set, at
selected bond distances between neighboring H atoms $R_\text{H--H}$ (in 
\AA).\protect\footnotemark[1]}
\begin{ruledtabular}
\begin{tabular*}{\textwidth}{@{\extracolsep{\fill}} l d d d d d d}
		\multirow{2}{*}{$R_\text{H--H}$} & 
		\multicolumn{1}{c}{\multirow{2}{*}{CCSD}} &
		\multicolumn{4}{c}{ACCSD} &
		\multicolumn{1}{c}{\multirow{2}{*}{FCI}} \\\cline{3-6}
		& &
		\multicolumn{1}{c}{(1,3)} &
		\multicolumn{1}{c}{$(1,\tfrac{3+4}{2})$\protect\footnotemark[2]} & 
		\multicolumn{1}{c}{$(1, 3 \times \tfrac{n_\text{o}}{n_\text{o} + 
				n_\text{u}} + 4 \times \tfrac{n_\text{u}}{n_\text{o} + 
				n_\text{u}})$} &
		\multicolumn{1}{c}{(1,4)} \\ [1mm]
\hline
		0.6 &   2.461 &  0.412 &  0.945 &  1.204 &  1.458 & -4.581177 \\
		0.7 &   2.942 &  0.498 &  1.071 &  1.348 &  1.619 & -5.133564 \\
		0.8 &   3.519 &  0.708 &  1.314 &  1.606 &  1.890 & -5.400721 \\
		0.9 &   4.157 &  1.053 &  1.689 &  1.992 &  2.285 & -5.513021 \\
		1.0 &   4.878 &  1.591 &  2.265 &  2.582 &  2.886 & -5.538852 \\
		1.1 &   5.691 &  2.435 &  3.164 &  3.501 &  3.821 & -5.516586 \\
		1.2 &   6.510 &  3.744 &  4.552 &  4.917 &  5.257 & -5.468944 \\
		1.3 &   7.013 &  5.701 &  6.620 &  7.022 &  7.388 & -5.409818 \\
		1.4 &   6.288 &  8.451 &  9.520 &  9.972 & 10.370 & -5.347723 \\
		1.5 &   1.896 & 11.987 & 13.258 & 13.773 & 14.209 & -5.287756 \\
		1.6 & -13.106 & 16.041 & 17.575 & 18.168 & 18.647 & -5.232798 \\
		1.7 & -66.276 & 20.069 & 21.930 & 22.618 & 23.144 & -5.184271 \\
		1.8 & \multicolumn{1}{c}{NC\protect\footnotemark[3]} 
		& 23.410 & 25.650 & 26.446 & 27.026 & -5.142644 \\
		1.9 & \multicolumn{1}{c}{NC\protect\footnotemark[3]} 
		& 25.531 & 28.163 & 29.075 & 29.715 & -5.107796 \\
		2.0 & \multicolumn{1}{c}{NC\protect\footnotemark[3]} 
		& 26.191 & 29.172 & 30.197 & 30.903 & -5.079254 \\
		2.1 & \multicolumn{1}{c}{NC\protect\footnotemark[3]} 
		& 25.462 & 28.691 & 29.809 & 30.585 & -5.056349 \\
		2.2 & \multicolumn{1}{c}{NC\protect\footnotemark[3]} 
		& 23.636 & 26.967 & 28.146 & 28.985 & -5.038308 \\
		2.3 & \multicolumn{1}{c}{NC\protect\footnotemark[3]} 
		& 21.090 & 24.372 & 25.568 & 26.453 & -5.024332 \\
		2.4 & \multicolumn{1}{c}{NC\protect\footnotemark[3]} 
		& 18.193 & 21.298 & 22.468 & 23.370 & -5.013655 \\
		2.5 & \multicolumn{1}{c}{NC\protect\footnotemark[3]} 
		& 15.248 & 18.087 & 19.193 & 20.080 & -5.005591 \\
\hline
		MUE\protect\footnotemark[4]
		& \multicolumn{1}{c}{NA\protect\footnotemark[5]} & 12.572 & 14.315 & 14.980 & 15.505 & \multicolumn{1}{c}{---} \\
		MSE\protect\footnotemark[6]
		& \multicolumn{1}{c}{NA\protect\footnotemark[5]} & 12.572 & 14.315 & 14.980 & 15.505 & \multicolumn{1}{c}{---} \\
\end{tabular*}
\end{ruledtabular}
\footnotetext[1]{\setlength{\baselineskip}{1em}
		The FCI energies are total energies in hartree, whereas all of the 
		remaining energies are errors relative to FCI in millihartree.}
\footnotetext[2]{\setlength{\baselineskip}{1em}
                Equivalent to the DCSD approach of Ref.\ \ocite{Kats2013a}.}
\footnotetext[3]{\setlength{\baselineskip}{1em}
		No convergence.}
\footnotetext[4]{\setlength{\baselineskip}{1em}
		Mean unsigned error.}
\footnotetext[5]{\setlength{\baselineskip}{1em}
		The mean errors are not reported because CCSD does not converge at larger values of $R_\text{H--H}$.}
\footnotetext[6]{\setlength{\baselineskip}{1em}
		Mean signed error.}
\end{table*}

\begin{table*}[h!]
\caption{\label{ACP_H10_active}
A comparison of the energies obtained with the CCSDt and various
ACCSDt approaches with the exact FCI
data for the symmetric dissociation of the $\text{H}_{10}$
ring, as described by the DZ basis set, at
selected bond distances between neighboring H atoms $R_\text{H--H}$ (in 
\AA).\protect\footnotemark[1]}
\begin{ruledtabular}
\begin{tabular*}{\textwidth}{@{\extracolsep{\fill}} l d d d d d d}
		\multirow{2}{*}{$R_\text{H--H}$} & 
		\multicolumn{1}{c}{\multirow{2}{*}{CCSDt}} &
		\multicolumn{4}{c}{ACCSDt} &
		\multicolumn{1}{c}{\multirow{2}{*}{FCI}} \\\cline{3-6}
		& &
		\multicolumn{1}{c}{(1,3)} &
		\multicolumn{1}{c}{$(1,\tfrac{3+4}{2})$} & 
		\multicolumn{1}{c}{$(1, 3 \times \tfrac{n_\text{o}}{n_\text{o} + 
				n_\text{u}} + 4 \times \tfrac{n_\text{u}}{n_\text{o} + 
				n_\text{u}})$} &
		\multicolumn{1}{c}{(1,4)} \\ [1mm]
\hline
		0.6 &   0.455 &  -1.751 &  -1.181 &  -0.904 &  -0.633 & -4.581177 \\
		0.7 &   0.554 &  -2.110 &  -1.502 &  -1.209 &  -0.922 & -5.133564 \\
		0.8 &   0.416 &  -2.726 &  -2.081 &  -1.773 &  -1.473 & -5.400721 \\
		0.9 &   0.349 &  -3.218 &  -2.541 &  -2.221 &  -1.912 & -5.513021 \\
		1.0 &   0.181 &  -3.735 &  -3.019 &  -2.686 &  -2.369 & -5.538852 \\
		1.1 &  -0.084 &  -4.169 &  -3.399 &  -3.050 &  -2.724 & -5.516586 \\
		1.2 &  -0.650 &  -4.477 &  -3.634 &  -3.266 &  -2.933 & -5.468944 \\
		1.3 &  -1.956 &  -4.591 &  -3.654 &  -3.266 &  -2.931 & -5.409818 \\
		1.4 &  -4.961 &  -4.457 &  -3.399 &  -2.994 &  -2.672 & -5.347723 \\
		1.5 & -11.879 &  -4.091 &  -2.881 &  -2.468 &  -2.182 & -5.287756 \\
		1.6 & -28.679 &  -3.630 &  -2.233 &  -1.824 &  -1.609 & -5.232798 \\
		1.7 & -81.643 &  -3.315 &  -1.692 &  -1.300 &  -1.193 & -5.184271 \\
		1.8 & \multicolumn{1}{c}{NC\protect\footnotemark[2]}
		&  -3.398 &  -1.510 &  -1.144 &  -1.178 & -5.142644 \\
		1.9  & \multicolumn{1}{c}{NC\protect\footnotemark[2]} 
		&  -4.043 &  -1.862 &  -1.523 &  -1.712 & -5.107796 \\
		2.0 & \multicolumn{1}{c}{NC\protect\footnotemark[2]} 
		&  -5.295 &  -2.810 &  -2.494 &  -2.832 & -5.079254 \\
		2.1 & \multicolumn{1}{c}{NC\protect\footnotemark[2]} 
		&  -7.106 &  -4.333 &  -4.029 &  -4.497 & -5.056349 \\
		2.2 & \multicolumn{1}{c}{NC\protect\footnotemark[2]} 
		&  -9.380 &  -6.355 &  -6.055 &  -6.626 & -5.038308 \\
		2.3 & \multicolumn{1}{c}{NC\protect\footnotemark[2]} 
		& -12.001 &  -8.762 &  -8.462 &  -9.114 & -5.024332 \\
		2.4 & \multicolumn{1}{c}{NC\protect\footnotemark[2]} 
		& -14.806 & -11.381 & -11.083 & -11.806 & -5.013655 \\
		2.5 & \multicolumn{1}{c}{NC\protect\footnotemark[2]} 
		& -17.551 & -13.966 & -13.660 & -14.471 & -5.005591 \\
\hline
		MUE\protect\footnotemark[3]
		& \multicolumn{1}{c}{NA\protect\footnotemark[4]} &   5.793 &   4.110 &   3.770 &   3.790 & \multicolumn{1}{c}{---} \\
		MSE\protect\footnotemark[5]
		& \multicolumn{1}{c}{NA\protect\footnotemark[4]} &  -5.793 &  -4.110 &  -3.770 &  -3.790 & \multicolumn{1}{c}{---} \\
\end{tabular*}
\end{ruledtabular}
	\footnotetext[1]{\setlength{\baselineskip}{1em}
		The FCI energies are total energies in hartree, whereas all of the 
		remaining energies are errors relative to FCI in millihartree. The
                CCSDt and ACCSDt approaches employed five active occupied and five
                active unoccupied MOs corresponding to the 1$s$ shells of the
                individual H atoms.}
	\footnotetext[2]{\setlength{\baselineskip}{1em}
		No convergence.}
	\footnotetext[3]{\setlength{\baselineskip}{1em}
		Mean unsigned error.}
	\footnotetext[4]{\setlength{\baselineskip}{1em}
                The mean errors are not reported because CCSDt does not converge at larger values of $R_\text{H--H}$.}
	\footnotetext[5]{\setlength{\baselineskip}{1em}
		Mean signed error.}
\end{table*}

\begin{table*}[h!]
\caption{\label{ACP_H10_fullt}
A comparison of the energies obtained with the CCSDT and various
ACCSDT approaches with the exact FCI
data for the symmetric dissociation of the $\text{H}_{10}$
ring, as described by the DZ basis set, at
selected bond distances between neighboring H atoms $R_\text{H--H}$ (in
\AA).\protect\footnotemark[1]}
\begin{ruledtabular}
\begin{tabular*}{\textwidth}{@{\extracolsep{\fill}} l d d d d d d}
		\multirow{2}{*}{$R_\text{H--H}$} & 
		\multicolumn{1}{c}{\multirow{2}{*}{CCSDT}} &
		\multicolumn{4}{c}{ACCSDT} &
		\multicolumn{1}{c}{\multirow{2}{*}{FCI}} \\\cline{3-6}
		& &
		\multicolumn{1}{c}{(1,3)} &
		\multicolumn{1}{c}{$(1,\tfrac{3+4}{2})$} & 
		\multicolumn{1}{c}{$(1, 3 \times \tfrac{n_\text{o}}{n_\text{o} + 
				n_\text{u}} + 4 \times \tfrac{n_\text{u}}{n_\text{o} + 
				n_\text{u}})$} &
		\multicolumn{1}{c}{(1,4)} \\ [1mm]
\hline
		0.6 &   0.065 &  -2.167 &  -1.591 &  -1.311 &  -1.037 & -4.581177 \\
		0.7 &   0.094 &  -2.602 &  -1.986 &  -1.688 &  -1.398 & -5.133564 \\
		0.8 &   0.110 &  -3.054 &  -2.404 &  -2.093 &  -1.790 & -5.400721 \\
		0.9 &   0.101 &  -3.485 &  -2.805 &  -2.482 &  -2.172 & -5.513021 \\
		1.0 &   0.035 &  -3.894 &  -3.175 &  -2.841 &  -2.523 & -5.538852 \\
		1.1 &  -0.170 &  -4.262 &  -3.491 &  -3.141 &  -2.815 & -5.516586 \\
		1.2 &  -0.703 &  -4.535 &  -3.691 &  -3.323 &  -2.989 & -5.468944 \\
		1.3 &  -1.991 &  -4.631 &  -3.693 &  -3.305 &  -2.969 & -5.409818 \\
		1.4 &  -4.986 &  -4.487 &  -3.428 &  -3.023 &  -2.700 & -5.347723 \\
		1.5 & -11.897 &  -4.115 &  -2.904 &  -2.490 &  -2.205 & -5.287756 \\
		1.6 & -28.687 &  -3.649 &  -2.251 &  -1.842 &  -1.626 & -5.232798 \\
		1.7 & -81.631 &  -3.330 &  -1.706 &  -1.313 &  -1.206 & -5.184271 \\
		1.8 & \multicolumn{1}{c}{NC\protect\footnotemark[2]} 
		&  -3.408 &  -1.520 &  -1.153 &  -1.187 & -5.142644 \\
		1.9 & \multicolumn{1}{c}{NC\protect\footnotemark[2]} 
		&  -4.049 &  -1.867 &  -1.528 &  -1.717 & -5.107796 \\
		2.0 & \multicolumn{1}{c}{NC\protect\footnotemark[2]} 
		&  -5.296 &  -2.811 &  -2.494 &  -2.833 & -5.079254 \\
		2.1 & \multicolumn{1}{c}{NC\protect\footnotemark[2]} 
		&  -7.101 &  -4.329 &  -4.025 &  -4.493 & -5.056349 \\
		2.2 & \multicolumn{1}{c}{NC\protect\footnotemark[2]} 
		&  -9.371 &  -6.346 &  -6.045 &  -6.618 & -5.038308 \\
		2.3 & \multicolumn{1}{c}{NC\protect\footnotemark[2]} 
		& -11.988 &  -8.750 &  -8.449 &  -9.100 & -5.024332 \\
		2.4 & \multicolumn{1}{c}{NC\protect\footnotemark[2]} 
		& -14.790 & -11.365 & -11.068 & -11.791 & -5.013655 \\
		2.5  & \multicolumn{1}{c}{NC\protect\footnotemark[2]} 
		& -17.533 & -13.927 & -13.644 & -14.459 & -5.005591 \\
\hline
		MUE\protect\footnotemark[3]
		& \multicolumn{1}{c}{NA\protect\footnotemark[4]} &   5.887 &   4.202 &   3.863 &   3.881 & \multicolumn{1}{c}{---} \\
		MSE\protect\footnotemark[5]
		& \multicolumn{1}{c}{NA\protect\footnotemark[4]} &  -5.887 &  -4.202 &  -3.863 &  -3.881 & \multicolumn{1}{c}{---} \\
\end{tabular*}
\end{ruledtabular}
	\footnotetext[1]{\setlength{\baselineskip}{1em}
		The FCI energies are total energies in hartree, whereas all of the 
		remaining energies are errors relative to FCI in millihartree.}
	\footnotetext[2]{\setlength{\baselineskip}{1em}
		No convergence.}
	\footnotetext[3]{\setlength{\baselineskip}{1em}
		Mean unsigned error.}
	\footnotetext[4]{\setlength{\baselineskip}{1em}
                The mean errors are not reported because CCSDT does not converge at larger values of $R_\text{H--H}$.}
	\footnotetext[5]{\setlength{\baselineskip}{1em}
		Mean signed error.}
\end{table*}

\begin{table*}[h!]
\caption{\label{ACP_H50}
A comparison of the energies obtained with the CCSD, {\DCSD} = DCSD, CCSDT, and {\DCSDT} approaches
with the nearly exact LDMRG(500) data for the symmetric dissociation of the
$\text{H}_{50}$ linear chain, as described by the STO-6G basis set, at selected bond distances 
between neighboring H atoms $R_\text{H--H}$ (in
bohr).\protect\footnotemark[1]}
\begin{ruledtabular}
\begin{tabular*}{\textwidth}{@{\extracolsep{\fill}} l d d d d d}
		$R_\text{H--H}$ & 
		\multicolumn{1}{c}{CCSD} & 
		\multicolumn{1}{c}{{\DCSD}} &
		\multicolumn{1}{c}{CCSDT} & 
		\multicolumn{1}{c}{{\DCSDT}} &
		\multicolumn{1}{c}{LDMRG(500)} \\ [1mm]
\hline
		1.0 & 11.90 &  6.30 &  0.27 &  -6.68 & -17.28407 \\
		1.2 & 16.28 &  9.12 &  0.29 &  -9.08 & -22.94765 \\
		1.4 & 20.99 & 12.78 & -0.10 & -11.37 & -25.59378 \\
		1.6 & 26.01 & 17.82 & -1.71 & -13.26 & -26.71944 \\
		1.8 & 31.00 & 24.87 & -6.96 & -14.62 & -27.03865 \\
		2.0 & 34.60 & 34.30 & \multicolumn{1}{c}{NC\protect\footnotemark[2]}
		& -15.73 & -26.92609 \\
		2.4 & \multicolumn{1}{c}{NC\protect\footnotemark[2]} &  59.23 & 
		\multicolumn{1}{c}{NC\protect\footnotemark[2]} & 
		-22.04\protect\footnotemark[3] & -26.16057 \\
		2.8 & \multicolumn{1}{c}{NC\protect\footnotemark[2]} &  86.24 & 
		\multicolumn{1}{c}{NC\protect\footnotemark[2]} & 
		-22.47\protect\footnotemark[4] & -25.27480 \\
		3.2 & \multicolumn{1}{c}{NC\protect\footnotemark[2]} & 106.45 & 
		\multicolumn{1}{c}{NC\protect\footnotemark[2]} & 
		-18.78\protect\footnotemark[4] & -24.56828 \\
		3.6 & \multicolumn{1}{c}{NC\protect\footnotemark[2]} & 113.43 & 
		\multicolumn{1}{c}{NC\protect\footnotemark[2]} &  
		-4.66\protect\footnotemark[4] & -24.10277 \\
\hline
		MUE\protect\footnotemark[5] & 
\multicolumn{1}{c}{NA\protect\footnotemark[6]} &  47.05 & 
\multicolumn{1}{c}{NA\protect\footnotemark[7]} &  13.87 &
		\multicolumn{1}{c}{---} \\
		MSE\protect\footnotemark[8] &
\multicolumn{1}{c}{NA\protect\footnotemark[6]} &  47.05 &
\multicolumn{1}{c}{NA\protect\footnotemark[7]} & -13.87 &
		\multicolumn{1}{c}{---} \\
\end{tabular*}
\end{ruledtabular}
\footnotetext[1]{\setlength{\baselineskip}{1em}
		The LDMRG(500) energies, taken from Ref.\ \ocite{Hachmann2006},
		are total energies in hartree. The remaining energies are errors 
		relative to LDMRG(500) in millihartree. In this case $n_{\rm o} = n_{\rm u}$, so that
                $\mbox{{\ACCSDX}} = \mbox{{\DCSD}} = {\rm DCSD}$ and
                $\mbox{{\ACCSDTX}} = \mbox{{\DCSDT}}$.}
\footnotetext[2]{\setlength{\baselineskip}{1em}
		No convergence.}
\footnotetext[3]{\setlength{\baselineskip}{1em}
		We were unable to converge this energy to an accuracy better than 1 $\text{mE}_\text{h}$.
		The reported value corresponds to the last CC iteration.}
\footnotetext[4]{\setlength{\baselineskip}{1em}
                We were unable to converge this energy to an accuracy better than 0.1 $\text{mE}_\text{h}$.
                The reported value corresponds to the last CC iteration.}
\footnotetext[5]{\setlength{\baselineskip}{1em}
		Mean unsigned error.}
\footnotetext[6]{\setlength{\baselineskip}{1em}
                The mean errors are not reported because CCSD does not converge at larger values of $R_\text{H--H}$.}
\footnotetext[7]{\setlength{\baselineskip}{1em}
                The mean errors are not reported because CCSDT does not converge at larger values of $R_\text{H--H}$.}
\footnotetext[8]{\setlength{\baselineskip}{1em}
		Mean signed error.}
\end{table*}

\pagebreak

\pagebreak

\end{document}